\documentclass[12pt]{article}
\pdfoutput=1 
\usepackage{amsmath,amssymb,amsthm}
\usepackage{bbm,latexsym}
\usepackage{graphicx}
\usepackage{rotating} 
\usepackage{amsfonts} 
\usepackage{xcolor}
\usepackage{multirow}
\usepackage[numbers,sort&compress]{natbib}
\usepackage{authblk}
\newcommand{\bone}{\mathbbm{1}}
\newcommand{\diag}{\mbox{diag}}
\newcommand{\sixt}{\underline{16}}
\newcommand{\one}{\underline{1}}
\newcommand{\ten}{\underline{10}}
\newcommand{\hc}{\mathrm{H.c.}}
\newcommand{\zed}{\mathbbm{Z}}

\begin{document}
\title{
\LARGE Lepton mixing from the hidden sector}
\setcounter{footnote}{2}
\author[1]{Patrick Otto Ludl\thanks{E-mail: patrick.ludl@mpi-hd.mpg.de}}
\author[1,2]{Alexei Yu.\ Smirnov\thanks{E-mail: smirnov@mpi-hd.mpg.de}}
\affil[1]{{\small Max-Planck-Institut f\"ur Kernphysik, Saupfercheckweg 1, D-69117 Heidelberg, Germany}}
\affil[2]{{\small International Centre for Theoretical Physics, Strada Costiera 11, I-34100 Trieste, Italy}}

\date{24 August 2015}

\maketitle

\begin{abstract}
Experimental results indicate a possible relation between
the lepton and quark mixing matrices of the form $U_\text{PMNS} \approx V_\text{CKM}^{\dagger} U_X$,
where $U_X$ is a matrix with special structure related to
the mechanism of neutrino mass generation.
We propose a framework which can realize such a relation.
The main ingredients of the framework are
the double seesaw mechanism, SO(10) Grand Unification and
a hidden sector of theory. The latter is composed of singlets (fermions and bosons) of
the GUT symmetry with masses between the GUT and Planck scale.
The interactions in this sector obey
certain symmetries $G_\text{hidden}$.
We explore the conditions under which symmetries $G_\text{hidden}$
can produce flavour structures in the visible sector.
Here the key elements are the basis-⁠fixing symmetry
and mediators which communicate information about properties
of the hidden sector to the visible one.
The interplay of SO(10) symmetry, basis-fixing symmetry identified as $\zed_2 \times \zed_2$
and $G_\text{hidden}$ can lead
to the required form of $U_X$. A different kind of new physics is responsible for
generation of the CKM mixing.
We present the simplest realizations of the framework which differ
by nature of the mediators and by symmetries of the hidden sector.
\end{abstract}

\section{Introduction}

There are various indications that in spite of their big difference the quark and
lepton mixings are somehow related. One appealing possibility can be formulated
as a relation between the lepton mixing matrix, $U_\text{PMNS}$, and the quark mixing
matrix, $V_\text{CKM}$, of the following form~\cite{Giunti-Tanimoto,Minakata-Smirnov,Xing,Harada,Antusch1,Antusch2,Farzan,Picariello}:
\begin{equation}
\label{quark-lepton-mixing}
U_\text{PMNS} = U_\text{CKM}^\dagger U_X,
\end{equation}
where
\begin{equation} 
U_\text{CKM} \sim V_\text{CKM} 
\label{appr}
\end{equation}
is a unitary matrix
which may coincide with the CKM mixing matrix or, in general,
has the same hierarchical structure as $V_\text{CKM}$
with expansion parameter $\lambda=\mathrm{sin}\,\theta_C$. 
We use here the notation $U_\text{CKM}$ to underline the close connection to (the same origin as) 
the quark mixing matrix $V_\text{CKM}$. 
The unitary matrix $U_X$ is related to additional structures
in the lepton sector which are responsible for the smallness of neutrino masses,
and it may be of special form originating from certain symmetries.
To be in agreement with the data, $U_X$  
should have vanishing (or small) 1-3 mixing and large (or even maximal) 2-3 mixing.

The relation (\ref{quark-lepton-mixing}) has been explored on pure 
phenomenological grounds in \cite{Giunti-Tanimoto}. It
has been proposed in the framework of 
quark-lepton complementarity \cite{Minakata-Smirnov} with 
$U_X = \Gamma_\alpha U_\text{BM}$, where $U_\text{BM}$ is the bimaximal mixing matrix \cite{bimaximal1,bimaximal2} and 
$\Gamma_\alpha = \mathrm{diag}(e^{i \alpha_e}, 1, 1)$. Later variations of 
(\ref{quark-lepton-mixing}) have been 
explored, in particular the TBM-Cabibbo mixing scheme \cite{King}
with $U_\text{CKM} = U_{12}(\theta_C)$ 
and $U_X = U_\text{TBM}$, where $U_\text{TBM}$ is the tri-bimaximal mixing
matrix~\cite{HPS-TBM}. Also
golden-ratio mixing~\cite{golden-ratio} has been considered with
$U_X = U_\text{GR}$~\cite{King}. All these cases have maximal 2-3 mixing, zero 1-3 mixing 
but differ by the values of 1-2 mixing.   

The measured value of $\theta_{13}$  supports the relation 
(\ref{quark-lepton-mixing}). Indeed, for $U_\text{CKM} = U_{12}(\theta_C)$ and 
$U_X = U_{23}^\text{max} U_{12}$  (where $U_{12}$ is arbitrary) the lepton mixing  
according to (\ref{quark-lepton-mixing}) becomes   
$U_\text{PMNS} =  U_{12}(\theta_C)^T U_{23}^\text{max} U_{12}$.
Reducing it to the standard parameterization form leads immediately to
\begin{equation}
\mathrm{sin}^2 \theta_{13} = \frac{1}{2} \mathrm{sin}^2 \theta_C. 
\label{pred}
\end{equation}
Here the coefficient $1/2$ originates from maximal 2-3 mixing,
\textit{i.e.}\ $\mathrm{sin}^2\theta_{23}^X = 1/2$. 
Eq.~(\ref{pred}) was in agreement with data in the first approximation. 
However, recent precise measurements of the leptonic 1-3 mixing angle~\cite{DoubleChooz,DayaBay,RENO} 
show a deviation from (\ref{pred}) by about $3\sigma$. 
Indeed, with Cabibbo mixing $\mathrm{sin}\,\theta_{C} = 0.22537 \pm 0.00061$~\cite{PDG}
we have 
\begin{equation}
\frac{1}{2} \sin^2\theta_{C} = 0.02540 \pm 0.00014, 
\label{cabibbo}
\end{equation}
whereas the most accurate value of
$\mathrm{sin}^2(2\theta_{13}) = 0.084 \pm 0.005$~\cite{DayaBay}
gives
\begin{equation}
\mathrm{sin}^2 \theta_{13}  = 0.0215 \pm 0.0013. 
\label{sin13m}
\end{equation}
Notice that the relative difference between the values in (\ref{cabibbo})
and (\ref{sin13m}), $\sim 0.18$, is of the order of the small 
elements of the CKM matrix, $2\lambda^2 \sim 0.1$, and can therefore 
be substantially reduced if the CKM corrections---due to the use of the complete $V_\text{CKM}$ in
Eq.~(\ref{quark-lepton-mixing})---are taken into account~\cite{Minakata-Smirnov}. 
The remaining difference can be due to non-maximal 2-3 mixing in $U_X$. It can
originate from some difference between $U_\text{CKM}$ and $V_\text{CKM}$ which, in turn, can be 
related to the difference of the masses of the charged leptons and down-type quarks.

Various data sets indicate that apart from 
the ``visible'' sector of theory, a ``hidden sector'' exists which is 
composed of singlets of the Standard Model (or GUT) gauge symmetry group.
The hidden sector can be responsible for the dark sector of the Universe
which includes particles of the dark matter,
fields needed for inflation and particles involved in the generation of the lepton and
baryon asymmetries of the Universe. 
Sterile neutrinos~\cite{Kopp} of different masses
(very light $\sim 10^{-3}\,\text{eV}$~\cite{deHolanda}; eV-scale, as indicated by
LSND~\cite{LSND}, MiniBooNE~\cite{MiniBooNE} as well as the reactor~\cite{reactor-anomaly} and
Gallium~\cite{Gallex1,Gallex2,SAGE1,SAGE2} anomalies;
keV-scale for warm dark matter~\cite{WDM}) can be manifestations of the hidden sector. 
Finally, the hidden sector could be responsible for the generation of small neutrino masses.

The hidden sector may not follow a generation structure 
and the number of new fermions as well as bosons can be bigger or even much bigger
than three. The hidden sector particles may have their own interactions 
including gauge (dark photons) and Yukawa interactions.
Moreover, the hidden sector may have its own symmetries $G_\text{hidden}$, 
but there can be also some common symmetries with the visible sector.
The origin of these symmetries as well as the components of the hidden sector
can be compactification of extra dimensions in string theory~\cite{Buchmueller}.

In this paper we will update on the relation (\ref{quark-lepton-mixing}). We argue that 
it suggests
the double seesaw mechanism, Grand Unification
and the presence of a hidden sector of theory.
We propose a framework in which the required form of the matrix $U_X$ 
originates from symmetries of the hidden sector, whereas
$V_\text{CKM}$ is generated by another kind of new physics.

The paper is organized as follows. In section~\ref{sec:mixing}
we consider the status of relation~(\ref{quark-lepton-mixing})
and discuss its implications. 
In section~\ref{framework} we formulate a framework which allows to 
realize the relation~(\ref{quark-lepton-mixing}). Here the main 
ingredients of the framework, and in particular the required symmetries, are considered. 
Several specific realizations are presented in section~\ref{realizations}. 
In section~\ref{additional_singlets} we consider effects of additional 
fermions from the hidden sector. Section~\ref{ckmphysics} is devoted to 
the new physics which is responsible for the CKM-type mixings in the lepton 
sector. Discussion and conclusions follow in section~\ref{discussion}.

\section{The relation between $U_\text{PMNS}$ and $V_\text{CKM}$
and its implications}
\label{sec:mixing}

\begin{figure}
\begin{center}
\includegraphics[height=7cm]{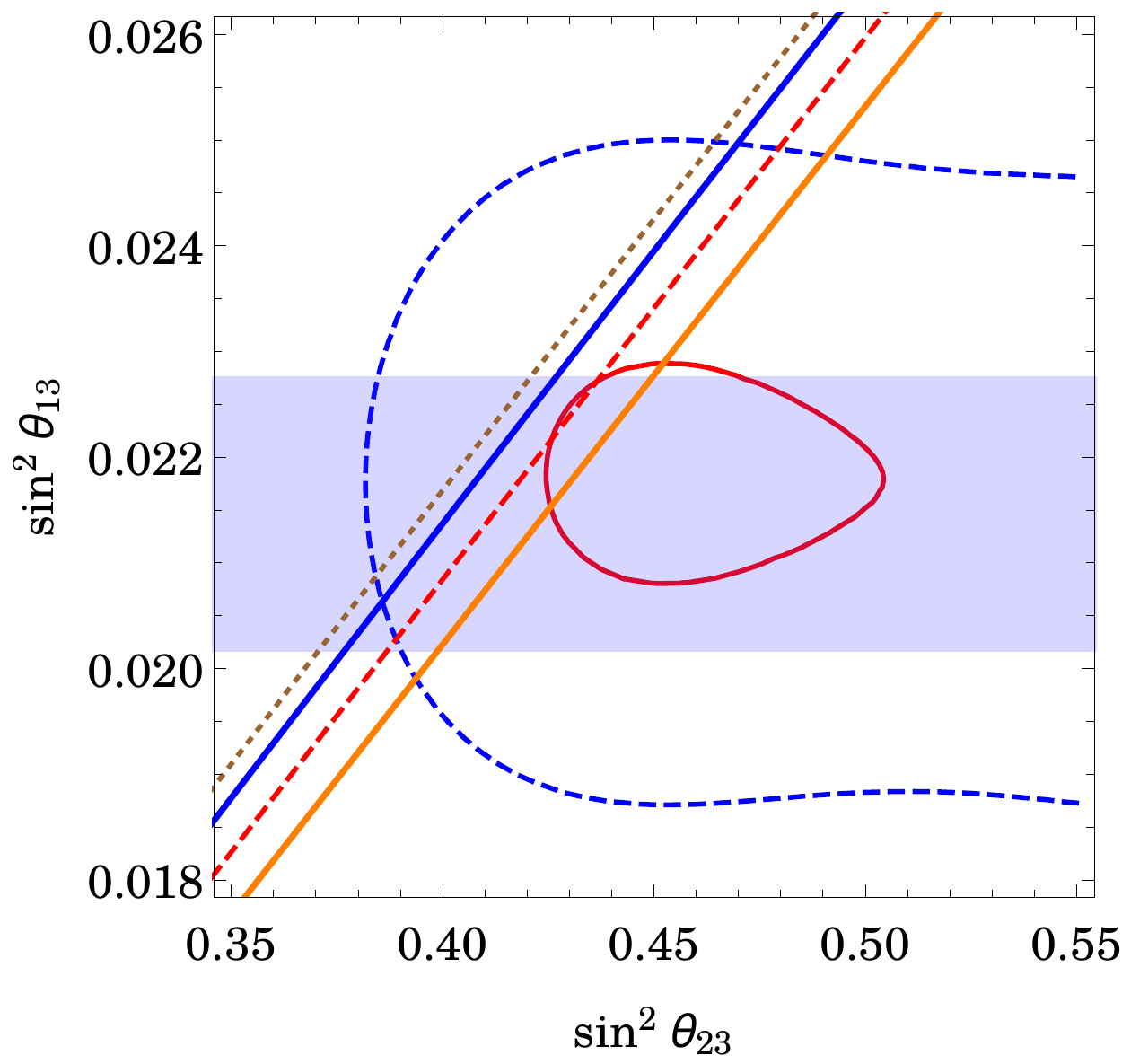}\hspace*{5mm}
\includegraphics[height=7cm]{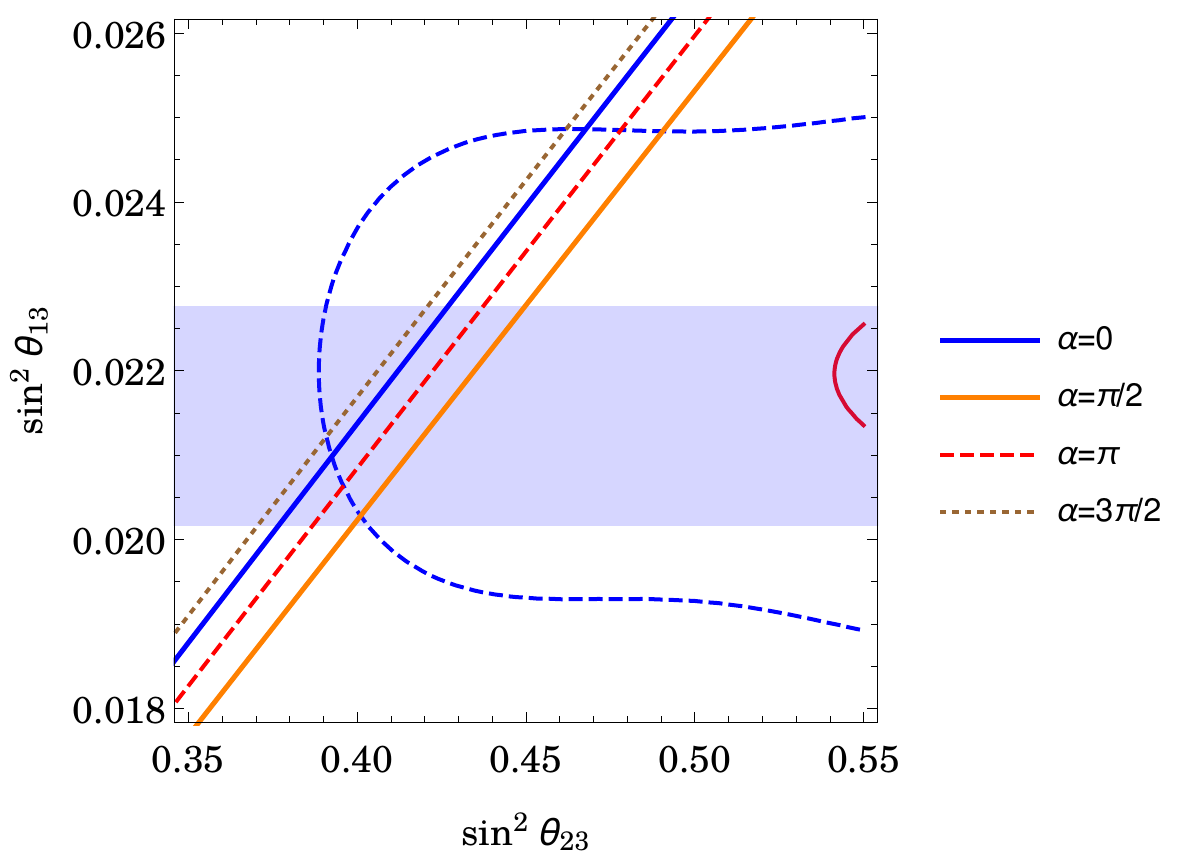}
\end{center}
\caption{Relation between $\mathrm{sin}^2\theta_{23}$ and $\mathrm{sin}^2\theta_{13}$
from Eq.~(\ref{quark-lepton-mixing}) with $\theta_{13}^X=0$
for different values of $\alpha$. The CKM parameters
have been set to the best-fit values of~\cite{PDG}.
The $1\sigma$ (red solid line) and $3\sigma$ (blue dashed line) regions according to
the global fit of~\cite{Schwetz,NuFIT} are shown for 
normal ordering (left plot) and inverted ordering (right plot). 
The blue band corresponds to the $1\sigma$-range for $\mathrm{sin}^2\theta_{13}$ allowed
by the recent results of the DayaBay experiment~\cite{DayaBay}.}
\label{figure1}
\end{figure}

\begin{figure}
\begin{center}
\includegraphics[height=7cm]{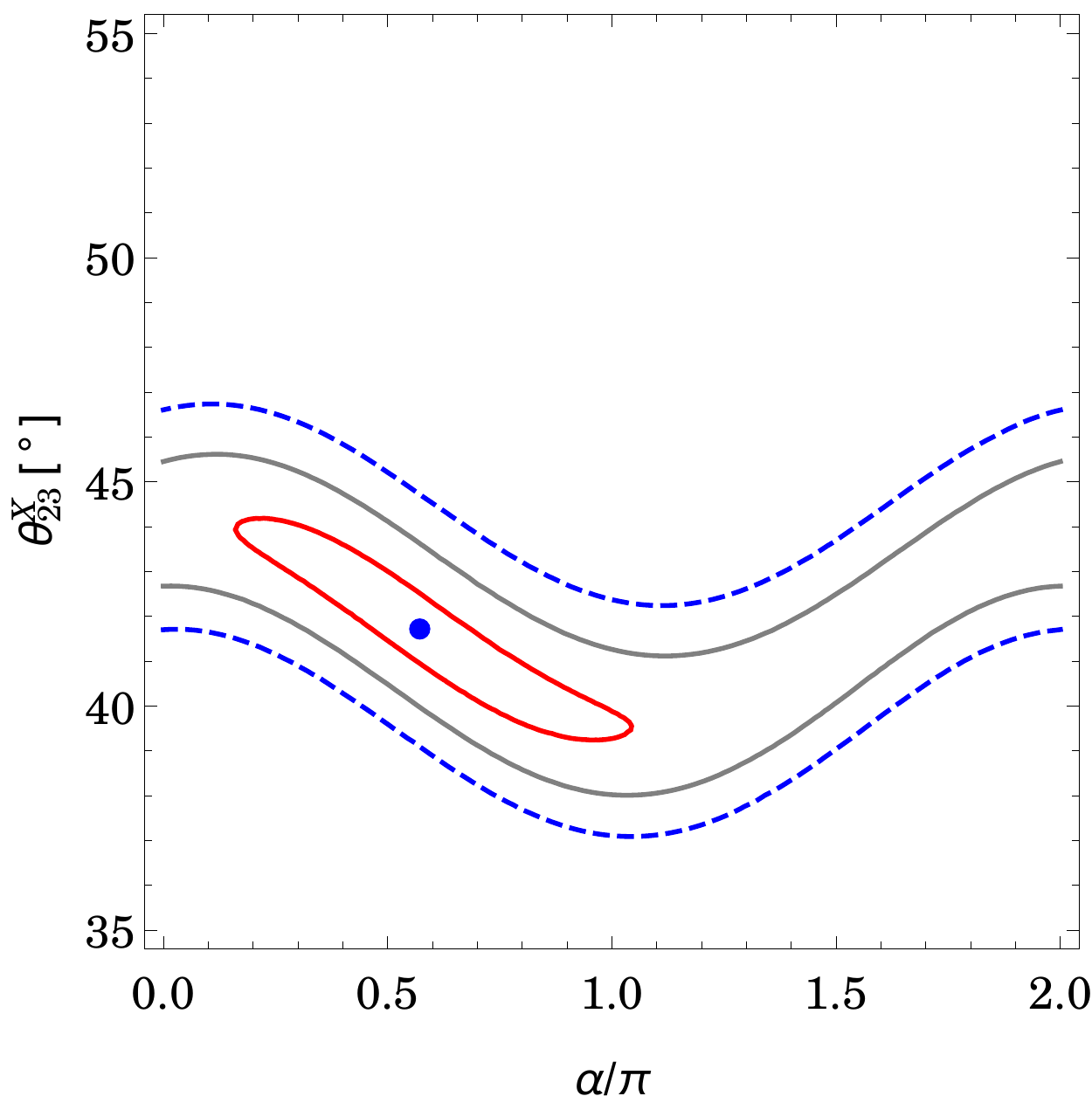}\hspace*{5mm}
\includegraphics[height=7cm]{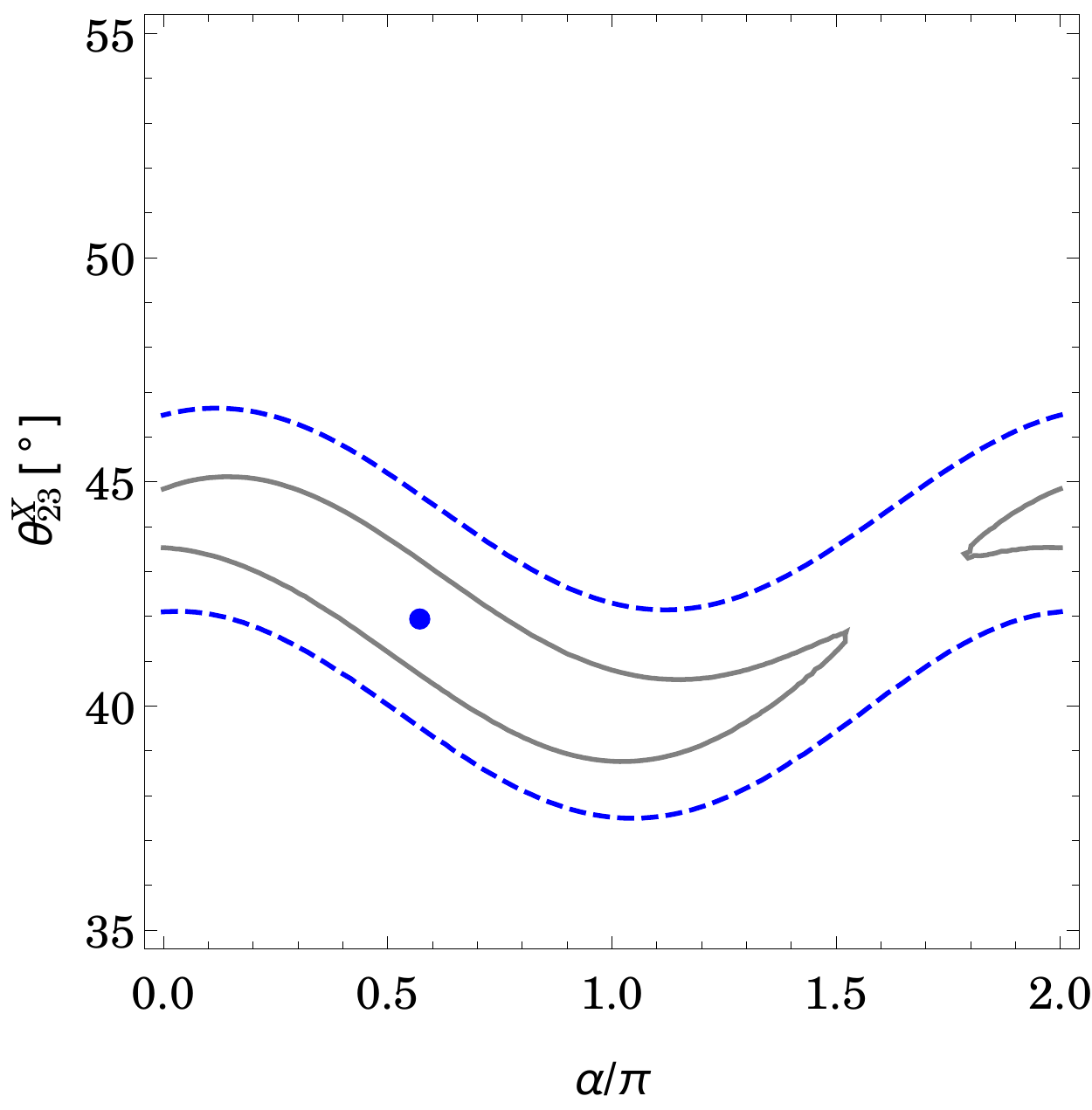}
\end{center}
\caption{The allowed regions for the parameters $\alpha$ and $\theta_{23}^X$
which reproduce relation~(\ref{quark-lepton-mixing}).
These regions have been computed from the
two-dimensional projection of the $\chi^2$-function of the global fit of~\cite{Schwetz,NuFIT}
into the $(\mathrm{sin}^2\theta_{23},\,\mathrm{sin}^2\theta_{13})$-plane.
The red solid, grey solid and blue dashed lines are the boundaries of the $1\sigma$, $2\sigma$
and $3\sigma$ regions, respectively. The blue dot corresponds to the
best fit point.
Left plot: normal mass ordering, 
right plot: inverted mass ordering.}
\label{figure2}
\end{figure}

Let us consider $U_X$ of general form  with the only restriction that  
the 1-3 mixing is vanishing or very small:  
\begin{equation}
U_X =  \Gamma\,U_{23}(\theta_{23}^X) \, U_{12}(\theta_{12}^X),\quad
\Gamma \equiv \mathrm{diag}(1,\,e^{i\varphi_2},\,e^{i\varphi_3}). 
\label{uxgen}
\end{equation}
Here we have omitted the Majorana phases of neutrinos and absorbed one overall phase of $\Gamma$
into the charged-lepton fields. 
Then, with (\ref{uxgen}) and exact equality $U_\text{CKM} = V_\text{CKM}$,
which we will use for definiteness,
the relation~(\ref{quark-lepton-mixing}) 
yields expressions for the mixing parameters
of $U_\text{PMNS}$ in terms of $\theta_{ij}^X$ and $\varphi_k$. Thus, the 1-3 mixing
equals
\begin{equation}
\label{s13sq}
 \mathrm{sin}^2\theta_{13}  = \lambda^2\, \mathrm{sin}^2\theta_{23}^X
  \left\{ 1 + 2 \frac{|V_{td}|}{V_{cd}}\, \mathrm{cot}\,\theta_{23}^X\,  
\mathrm{cos}(\alpha + \mathrm{Arg}V_{td}) + \mathcal{O}(\lambda^4) \right\} , 
\end{equation}
where 
\begin{equation}
\alpha \equiv \varphi_2 - \varphi_3 
\end{equation}
and $\mathrm{Arg}V_{td} = - 21.8^{\circ} = -0.12\pi$.
Using the Wolfenstein parameterization~\cite{Wolfenstein} for $V_\text{CKM}$,
Eq.~(\ref{s13sq}) can be rewritten as 
\begin{equation}
\mathrm{sin}^2\theta_{13} = \lambda^2\, \mathrm{sin}^2\theta_{23}^X
  \left\{ 1 - 2 A \lambda^2 \sqrt{(1-\rho^2)+\eta^2} \,
\mathrm{cot}\,\theta_{23}^X\, \mathrm{cos}(\alpha + \mathrm{Arg}V_{td}) \right\} 
+ \mathcal{O}(\lambda^6),
  \end{equation}
where $\mathrm{Arg}V_{td} = \mathrm{arctan}\frac{\eta}{\rho-1} + \mathcal{O}(\lambda^2)$.
For the 2-3 mixing we obtain 
\begin{equation}
\label{s23sq}
\mathrm{tan}^2\theta_{23}  = \mathrm{tan}^2\theta_{23}^X \, (1-\lambda^2)
\left\{
1 - \frac{4A\lambda^2\, \mathrm{cos}\,\alpha}{\mathrm{sin}\,2\theta_{23}^X}
\right\} + \mathcal{O}(\lambda^4). 
\end{equation}
Eliminating $\theta_{23}^X$ from Eqs.~(\ref{s13sq}) and~(\ref{s23sq}) immediately yields 
a relation between the lepton mixing parameters $\mathrm{sin}^2\theta_{13}$ 
and $\mathrm{sin}^2\theta_{23}$ as a function
of $\alpha$ (see Fig.~\ref{figure1}). Approximate analytic expressions for this relation can be 
obtained taking into account that near maximum $\mathrm{sin}\,2\theta_{23}^X$ only weakly depends on $\theta_{23}^X$.
Using $\mathrm{sin}\,2\theta_{23}^X=1$ in the denominator of~(\ref{s23sq}) we find
\begin{equation} 
\mathrm{sin}^2\theta_{13} \approx \lambda^2\, \frac{\mathrm{tan}^2\theta_{23} }{\zeta^2 +     
\mathrm{tan}^2\theta_{23}}
\left\{ 1 + 2 \frac{|V_{td}|}{V_{cd}}\, \zeta \mathrm{cot}\,\theta_{23}\,
\mathrm{cos}(\alpha + \mathrm{Arg}V_{td}) + \mathcal{O}(\lambda^4) \right\},
\end{equation}
where 
\begin{equation}
\zeta^2 (\alpha) \equiv  (1-\lambda^2)
(1 - 4A\lambda^2\, \mathrm{cos}\,\alpha )
\label{zeta}
\end{equation}
and we have used $\mathrm{tan}^2\theta_{23} \approx \zeta^2 \mathrm{tan}^2\theta_{23}^X$.
Fig.~\ref{figure2} shows the values of $\theta_{23}^X$ and $\alpha$
allowed by  experimental data.
Notice that for $\theta_{23}^X = \pi/4$ (exactly maximal mixing) and 
$\alpha = -\mathrm{Arg}\,V_{td}$ we have $\mathrm{sin}^2\theta_{23} = 0.45$ and
$\mathrm{sin}^2\theta_{13} = 0.0234$, which is just $1.5 \sigma$ above 
the best value from experiment.
%
%
%
%
%
%
%
%
%
%
%
As can be seen from Fig.~\ref{figure1} and Fig.~\ref{figure2},
the relation~(\ref{quark-lepton-mixing})
is in good agreement with experiment, especially for normal
mass ordering and $\alpha \lesssim \pi$.
Notice that these results should merely be considered as some orientation  
since in general $U_\text{CKM}$ can deviate from $V_\text{CKM}$. 
Still the coincidence even at the correction level looks very appealing 
and we assume that it is not accidental.

Finally, the 1-2 mixing is determined by 
\begin{equation}
\label{s12sq}
\mathrm{sin}^2\theta_{12}  = \mathrm{sin}^2\theta_{12}^X - \lambda\, \mathrm{sin}\,2\theta_{12}^X\,
\, \mathrm{cos}\,\theta_{23}^X\,\mathrm{cos}\,\varphi_2
 + \lambda^2\, \mathrm{cos}\,2\theta_{12}^X\, \mathrm{cos}^2\theta_{23}^X
 + \mathcal{O}(\lambda^3). 
\end{equation}
Fixing $\mathrm{sin}^2\theta_{12}$ to the best fit value of~\cite{Schwetz},
and using the best fit value $\theta_{23}^X \approx 42^\circ$ (see Fig.~\ref{figure2}) 
we find from (\ref{s12sq}) the required range $\mathrm{sin}^2\theta_{12}^X \in [0.16,\,0.47]$.

The relation (\ref{quark-lepton-mixing}) means that information about the quark mixing 
is communicated somehow to the lepton mixing. In turn, this implies 
a kind of quark-lepton unification or/and common flavour symmetries 
in the quark and lepton sector~\cite{Raidal}.
Furthermore, Eq.~(\ref{quark-lepton-mixing}) points towards
the type-I seesaw mechanism or its extensions~\cite{Seesaw1,Seesaw2,Seesaw3,Seesaw4,Seesaw5}.
Indeed, the Dirac neutrino mass matrix $m_D$ can be written as
\begin{equation}
 m_D = U_L \hat{m}_D U_R^\dagger,\quad \hat{m}_D \equiv  \diag (m_1^D, m_2^D, m_3^D) 
\end{equation}
with $U_L$ and $U_R$ being unitary matrices of transformations of the left- and right-handed neutrino components,
respectively.   
Then according to the seesaw mechanism the light-neutrino mass matrix is given by
\begin{equation}\label{mnuMX}
 m_\nu = -m_D M_R^{-1} m_D^T = U_L M_X U_L^T, 
\end{equation}
where 
\begin{equation}
M_X \equiv - \hat{m}_D U_R^\dagger M_R^{-1} U_R^\ast \hat{m}_D.
\end{equation}
If $M_X$ is diagonalized by a unitary matrix $U_X$, \textit{i.e.}\
\begin{equation}
U_X^T M_X U_X = \hat{M}_X,
\end{equation}
the light-neutrino mass matrix $m_\nu$ is diagonalized by
\begin{equation}
U_\nu = U_L^\ast U_X.
\end{equation}
The matrix $U_L$ can be related 
to the quark mixing matrix $V_\text{CKM}$
in grand unified theories~\cite{GUT1,GUT2}.
An immediate realization is an SO(10)-GUT~\cite{SO10-1,SO10-2} with a dominant 
contribution of Higgs ten-plet fields $\ten_H$
to the fermion mass terms. In this case all mass matrices are symmetric
with $m_D \propto m_u$ and $m_\ell \propto m_d$
and thus, in the basis where $m_\ell\propto m_d$ is diagonal,
\begin{equation}
V_\text{CKM} = U_u^\dagger = U_L^T 
\quad\Rightarrow\quad U_\text{PMNS} = U_\nu = V_\text{CKM}^\dagger U_X,
\end{equation}
\textit{i.e.}\ Eq.~(\ref{quark-lepton-mixing}).
%
For conditions allowing to realize the less restrictive relation
$\theta_{13} \approx \theta_{C}/\sqrt{2}$---see
equation~(\ref{pred})---in Pati-Salam and SU(5)-GUTs we refer the reader to~\cite{Sluka}.

According to our previous considerations,
$M_X$ should lead to vanishing or very small 1-3 mixing and close to maximal
2-3 mixing in $U_X$, \textit{i.e.}\ $M_X$ should be approximately invariant under the 2-3-permutation symmetry
($\mu\tau$-symmetry).
Since we also need a sizeable $\mathrm{sin}^2\theta_{12}^X \gtrsim 0.16$, the matrix $M_X$
should be close to the tri-bimaximal mass matrix
\begin{equation}
M_X \sim M_\text{TBM}.
\end{equation}
Furthermore, the light neutrinos have the weakest hierarchy among all known fermion species.
Therefore, also $M_X$ cannot be strongly hierarchical---see Eq.~(\ref{mnuMX}).
On the other hand, the mentioned SO(10)-scenario, or generically the assumption of
quark-lepton similarity $m_D \sim m_q \sim m_\ell$, suggests
a strong hierarchy of $m_D$. From this it follows that
\begin{equation}
M_R = -U_R^\ast \hat{m}_D M_\text{TBM}^{-1} \hat{m}_D U_R^\dagger
\end{equation}
is extremely hierarchical (quadratical in the up-type quark mass hierarchy).
This indicates that $M_R$ itself is generated by a
type-I seesaw mechanism, \textit{i.e.}\
the double seesaw mechanism~\cite{double-seesaw1,double-seesaw2} for $m_\nu$.

In order to implement the double seesaw, we add three heavy gauge-singlets $S$ 
to the fermion sector,
in which case the neutrino mass term reads
\begin{equation}\label{nu-massterm}
\mathcal{L}_{\nu}^\text{mass} = -\frac{1}{2}\overline{n_L} \mathcal{M} n_L^c +\hc,
\end{equation}
where
\begin{equation}
n_L=
\begin{pmatrix}
\nu_L\\ \nu_R^c\\ S
\end{pmatrix}
\end{equation}
and
\begin{equation}
\label{massmat}
\mathcal{M} =
\begin{pmatrix}
0 & m_D & m_{\nu S}\\
m_D^T & 0 & M_{RS}\\
m_{\nu S}^T & M_{RS}^T & M_{S}
\end{pmatrix}.
\end{equation}
Here $M_S$ is the Majorana mass matrix of the new heavy fermions $S$
and $M_{RS}$ is a Dirac-type neutrino mass matrix of  $\nu_R$ and $S$.
In general also the mass matrix $m_{\nu S}$ connecting $\nu_L$ with $S$ will
be present. If $M_S$ is invertible,
the right-handed Majorana neutrino mass matrix $M_R$ has the form
\begin{equation}
\label{MReff}
M_R \approx -M_{RS} M_{S}^{-1} M_{RS}^T,
\end{equation}
so if $M_{RS}$ is hierarchical, $M_R$ will have the desired
strong hierarchy.
The light-neutrino mass matrix $m_\nu$
is approximately given by
\begin{equation}\label{seesaw}
 m_\nu \approx m_\nu^{DS} + m_\nu^{LS},
\end{equation}
where
\begin{equation}\label{DS}
 m_\nu^{DS} = m_D( M_{RS}^{-1 T} M_{S} M_{RS}^{-1} ) m_D^T
\end{equation}
is the double seesaw contribution and
\begin{equation}\label{LS}
 m_\nu^{LS} = - \left[ m_D (m_{\nu S} M_{RS}^{-1})^T + (m_{\nu S} M_{RS}^{-1}) m_D^T \right]
\end{equation}
is the linear seesaw~\cite{linear-seesaw} contribution to $m_\nu$.
If $M_S$ is singular but $M_{RS}$ has rank three (and is thus invertible), we find
\begin{equation}
\begin{pmatrix}
 0 & M_{RS}\\
 M_{RS}^T & M_{S}
\end{pmatrix}^{-1} =
\begin{pmatrix}
 -(M_{RS}^{-1})^T M_S M_{RS}^{-1} & (M_{RS}^{-1})^T\\
 M_{RS}^{-1} & 0
\end{pmatrix}
\end{equation}
in which case Eqs.~(\ref{DS}) and~(\ref{LS}) still hold, while
Eq.~(\ref{MReff}) does not.

If $M_{RS}$ is at GUT-scale,
and the new fermions $S$ have masses at about one order of
magnitude below the Planck scale we obtain
$M_R \sim 10\, M_\text{GUT}^2 / M_\text{Pl} \sim 10^{14}\,\text{GeV}$.
From the study of the ranges of the masses of the right-handed
neutrinos in the case of $m_D \sim m_u$~\cite{Akhmedov}
one can find that, depending on the values of the
Dirac and Majorana phases of $m_\nu$, a value of the mass of the heaviest
right-handed neutrino $M_{R3} \sim 10^{14}\,\text{GeV}$ is possible
for a smallest neutrino mass $m_0\gtrsim (10^{-3} \div 10^{-2})\,\text{eV}$.
The mass ranges for the other two right-handed neutrinos 
in this case are $M_{R2} \sim (10^8 \div 10^{10})\,\text{GeV}$
and $M_{R1} \sim (10^4 \div 10^{8})\,\text{GeV}$~\cite{Akhmedov}.
If $m_0 \simeq (10^{-4} \div 10^{-3})\,\text{eV}$,  
a higher scale $M_{R3} \simeq (10^{15} \div 10^{16})\,\text{GeV}$
is required. 
The relevant mass scales for the double seesaw scenario
with $M_{R3} \sim 10^{14}\,\text{GeV}$ are shown in table~\ref{scales}.
\begin{table}
\begin{center}
\renewcommand{\arraystretch}{1.2}
\begin{tabular}{|l|l|}
\hline
Mass matrix & scale\\
\hline
$M_S$ & $\sim 10^{18}\,\text{GeV} \sim \frac{1}{10}M_\text{Pl}$\\
$M_{RS}$ & $\sim 10^{16}\,\text{GeV} \sim M_\text{GUT}$\\
$M_{R} \approx -M_{RS} M_{S}^{-1} M_{RS}^T$ & $\sim 10^{14}\,\text{GeV}$\\
\hline
$m_{\nu S}$ & $\sim 10^2\,\text{GeV} \sim M_\text{EW}$\\
$m_{D}$ & $\sim 10^2\,\text{GeV} \sim M_\text{EW}$\\
\hline
$m_\nu^{DS} = m_D( M_{RS}^{-1 T} M_{S} M_{RS}^{-1} ) m_D^T$ & $\sim 10^{-1}\,\text{eV}$\\
$m_\nu^{LS} = - \left[ m_D (m_{\nu S} M_{RS}^{-1})^T + (m_{\nu S} M_{RS}^{-1}) m_D^T \right]$ 
& $\sim 10^{-3}\,\text{eV}$\\
\hline
\end{tabular}
\renewcommand{\arraystretch}{1.0}
\caption{The different scales involved in the presented double seesaw framework.}\label{scales}
\end{center}
\end{table}
The linear seesaw term is dominated by the double seesaw contribution, but it may still
play a subleading role in the phenomenology of $m_\nu$.

To summarize, the realization of the relation~(\ref{quark-lepton-mixing}) implies
\begin{itemize}
 \item The seesaw mechanism,
 \item quark-lepton unification (\textit{e.g.}\ an SO(10)-GUT) or/and common flavour symmetries
in the quark and lepton sector,
 \item ``CKM physics'' leading to small quark mixing $V_\text{CKM}$, and
 \item new physics in the neutrino sector generating $U_X \sim U_\text{TBM}$.
\end{itemize}
Moreover, quark-lepton similarity indicates the double seesaw mechanism for
the generation of the light-neutrino mass matrix $m_\nu$. The new 
fermionic singlets can be components of the hidden sector of theory.

\section{Framework}
\label{framework}

The main ingredients of the framework which can realize
relation~(\ref{quark-lepton-mixing}) include
\begin{itemize}
 \item[(i)] SO(10) Grand Unification (although other GUT symmetries can be considered).
 \item[(ii)] The existence of a hidden sector composed of fermions and bosons,
 which are singlets of SO(10). The interactions in the hidden sector may have certain symmetries.
 \item[(iii)] The basis-fixing symmetry and mediators
which communicate information about the structure/interactions of the hidden sector to the visible one.
 \item[(iv)] The double (or even more complicated) seesaw mechanism which ensures 
complete or partial screening of the Dirac structures.
 \item[(v)] Separation of the physics responsible for the 
CKM mixing from the physics responsible for large neutrino mixing.
\end{itemize}
In the following we will discuss these ingredients in detail.

\subsection{The visible and the hidden sector}

%
We consider an SO(10) GUT with three families of fermions in 16-plets $\sixt_F$.  
The dominant contribution to the fermion mass terms is generated by ten-plet 
Higgs-fields $\ten_{H}$. Two or more $\ten_{H}$ are needed to generate the different mass hierarchies
of the up- and down-components of the doublets:
Namely, one Higgs ten-plet field $\ten_{H}^u$ gives rise to the up-type quark mass matrix $m_u$
and the Dirac neutrino mass matrix $m_D$, and another Higgs field $\ten_{H}^d$
is responsible for the mass matrices of the
down-type quarks $m_d$ and charged leptons $m_\ell$ and for CKM mixing~\cite{ten-u-d-1,ten-u-d-2}.
Additional physics is required
to generate the mass hierarchy of quarks and leptons.
Further complication is needed to explain the difference of the masses of the charged leptons and down-type quarks. 
We refer to all this as ``CKM new physics'' which we will comment on in section~\ref{ckmphysics}.  

%
In the following we will call the set of particles which have
non-trivial transformation properties under SO(10)
the ``visible sector'' of theory.
We refer to the hidden sector as to the
system of particles (fermions and bosons) and fields
which are singlets of SO(10). 
The interactions in this sector (Yukawa and new gauge interactions) 
may have a certain symmetry $G_\text{hidden}$.
The idea is that this hidden sector symmetry is responsible 
for the generation of $U_X$ with the required properties.
In general, the hidden symmetry can include several different factors and 
the hidden sector fields may have all possible charge assignments with respect to these factors.
Also, there can be 
some common symmetry in the hidden and visible sector and 
the charges of multiplets in the visible sector can be such 
that they allow to couple them with only few components from the 
hidden sector.  

For the remainder of this paper, the most important part of the
hidden sector will be gauge singlet fermions $S$, which are needed
in order to implement the double seesaw mechanism. However, also scalar gauge singlets $\one_\chi$
will play an important role for the realization of the hidden sector symmetry.
A priori, we may add an arbitrary number of SO(10)-singlet fermions $S$ with a
mass scale of $M_S \sim 10^{18}\,\text{GeV}$ to the fermion content
of our framework, which in the Lagrangian are denoted by $\one_S$. 
However, if there are less than three of these singlets coupling
to the active neutrinos,\footnote{Since we have an SO(10)-GUT in mind,
the term ``active neutrinos'' also includes $\nu_R$.} the matrix $M_{RS}$ will have rank smaller
than three, in which case the right-handed neutrino mass matrix
$M_R$ will be singular, a case we do not want to study here. 
Therefore, at least three singlets $S$ contributing to $M_{RS}$ should be introduced.
In what follows we will consider only three singlets which couple with the
visible sector directly. This is also needed to realize screening~\cite{screening1,screening2}
of the Dirac structures.
The case of more than three singlets will be considered in section~\ref{additional_singlets}.

To connect the visible and the hidden sector and generate $M_{RS}$ we need to introduce
scalar 16-plet(s) $\overline{\sixt}_H$.
The generation of the neutrino masses via the double seesaw mechanism allows us to avoid 
introduction of the high-dimensional multiplets 
$\underline{120}_H$ and $\underline{126}_H$.  
The absence of $\underline{120}_H$ and $\underline{126}_H$ is in fact desirable, because
including such high-dimensional scalar representations is known to give rise
to Landau poles in the gauge coupling already before reaching the Planck scale
$M_\text{Pl}\sim 10^{19}\,\text{GeV}$.

\subsection{Yukawa interactions, the neutrino portal and screening}

There are three types of Yukawa couplings in our framework. Their graphical representations
are shown in Fig.~\ref{mediators}.
\begin{enumerate}
 \item The visible sector couplings: 
\begin{equation} 
\mathcal{L}^{(FF)} = - Y^{(FF)}_{ab\alpha} \sixt_{Fa} \sixt_{Fb} \ten_{H}^\alpha + \hc,
\label{Yuk1}
\end{equation}
($a, b = 1,2,3$, $\alpha = u, d$)
which generate the Dirac mass matrices of fermions. Although the number of
Higgs ten-plets $\ten_{H}^u$ and $\ten_{H}^d$ could be arbitrary, we
will here consider only one $\ten_{H}^u$ and one $\ten_{H}^d$.
These interactions are also responsible for quark mixing (see section~\ref{ckmphysics}).
 \item The ``portal interactions''
\begin{equation}
\mathcal{L}^{(FS)} = - Y^{(FS)}_{ajk} \sixt_{Fa} \one_{Sj} \overline{\sixt}_{Hk} + \hc,
\label{Yuk2}
\end{equation}
which couple fermions of the visible and hidden sector
and thus provide the (neutrino) ``portal'' between the two sectors.
 \item The hidden sector interactions
\begin{equation}
\mathcal{L}^{(SS)} = - \frac{1}{2} Y^{(SS)}_{ijk} \one_{Si} \one_{Sj} \one_{\chi k} + \hc, 
\label{Yuk3} 
\end{equation}
where $\one_{\chi k}$ are scalar SO(10)-singlets.
These interactions include only particles of the hidden sector.  
\end{enumerate}
The neutrino mass term is given by Eqs.~(\ref{nu-massterm})-(\ref{massmat}) with the mass matrices
\begin{equation}
m_D = Y^{(FF)}_u \langle \ten_H^u \rangle,\quad
m_{\nu S} = Y^{(FS)} \langle \overline{\sixt}_{H} \rangle,\quad
 M_{RS} = Y^{(FS)} \langle \overline{\sixt}_{H} \rangle,\quad
 M_S = Y^{(SS)} \langle \one_\chi \rangle.
\end{equation}
Note that $m_{\nu S}$ and $M_{RS}$ are generated by different components
of the VEVs of the $\overline{\sixt}_H$.
If the VEVs of $\overline{\sixt}_H$ are
at about GUT scale, the mass of the heaviest right-handed neutrino is given
by $M_{R3} \sim 10^{14}\,\text{GeV}$.
Since we have not introduced $\underline{126}_H$, there is no
right-handed neutrino mass term.\footnote{The possible dimension-5 contribution
to $M_R$ stemming from $(1/\Lambda)\sixt_F\sixt_F\overline{\sixt}_H\overline{\sixt}_H$ can
easily be forbidden by a discrete symmetry.}

Let us introduce the matrix
\begin{equation}
D \equiv m_D (M_{RS}^{-1})^T
\end{equation}
so that $m_\nu^{DS} = D M_S D^T$.
Then the simplest way to obtain the correct hierarchical structures of $M_R$ and $m_\nu$
is to generate a mild hierarchy in the matrices $M_S$ and $D$.
For this the strong hierarchies of $m_D$ and $M_{RS}$ should at least partially
cancel each other in $D$, and consequently the light-neutrino mass matrix $m_\nu$ will have a hierarchy
similar to $M_S$. If $m_D \propto M_{RS}$, we have $D \propto \mathbbm{1}$, \textit{i.e.}\ complete
\textit{screening}~\cite{screening1,screening2} of the Dirac neutrino mass matrix.
In this case $m_\nu^{DS} \propto M_{S}$ and the difference between
$U_\text{PMNS}$ and $V_\text{CKM}^\dagger$ directly reflects the structure of the
mass matrix in the hidden sector.
In the following we will show how screening and large neutrino mixing from the hidden
sector can be obtained in our framework using symmetries in the visible
and the hidden sector.

\subsection{Basis-fixing symmetry and mediators}
\label{section-mediators}

In this section we formulate conditions under which  
symmetries of the hidden sector can affect the flavour structure of the visible sector,
and eventually lead to the required form of $U_X$.

In order to
assure that the hidden sector symmetries can influence the form of $m_\nu$,
we must guarantee ``communication'' between the two sectors,
which happens in the portal interaction $\mathcal{L}^{(FS)}$.
In general the portal interaction is a sum over operators of the
form
\begin{equation}
O_\text{vis} \times O_\text{hidden},
\label{portal}
\end{equation}
where $O_\text{vis}$ and $O_\text{hidden}$ are operators
containing only visible and hidden sector fields, respectively.
If there are no visible sector fields
transforming under $G_\text{hidden}$ or hidden sector fields transforming
under $G_\text{vis}$ (the symmetry of the visible sector), then
\begin{equation}
\mathcal{L}^{(FS)} = \sum_{j,k} C_{jk}\, O_\text{vis}^j \times O_\text{hidden}^k + \hc
\label{portal2}
\end{equation}
and the coefficients $C_{jk}$ in front of 
the products of invariants
$O_\text{vis}^j$ of $G_\text{vis}$ and $O_\text{hidden}^k$ of $G_\text{hidden}$
are unrestricted by both $G_\text{vis}$
and $G_\text{hidden}$, and are thus free parameters of the theory.
Consequently, there are no restrictions of the
hidden sector symmetry $G_\text{hidden}$ on the flavour structure of the visible sector.
In order to have communication of information of the hidden sector
to the visible sector, $\mathcal{L}^{(FS)}$ must not factorize as in Eq.~(\ref{portal2}).
Therefore, some symmetry---$G_\text{basis}$---should exist which
acts both in the visible and the hidden sector.

$G_\text{basis}$ should at least fix a basis in both sectors
and we call it the basis-fixing symmetry.
We will call the fields, which in this case provide the communication
between the two sectors, the mediators.
In order to fix a basis, $G_\text{basis}$ must be a symmetry
which can differentiate among the three generations. The smallest potential candidates for
$G_\text{basis}$ are therefore $\zed_3$ and $\zed_2\times\zed_2$. In our examples we will
always use $\zed_2\times\zed_2$ which also makes $\mathcal{L}^{(FF)}$ diagonal.

\begin{figure}
\begin{center}
\includegraphics[width=0.6\textwidth]{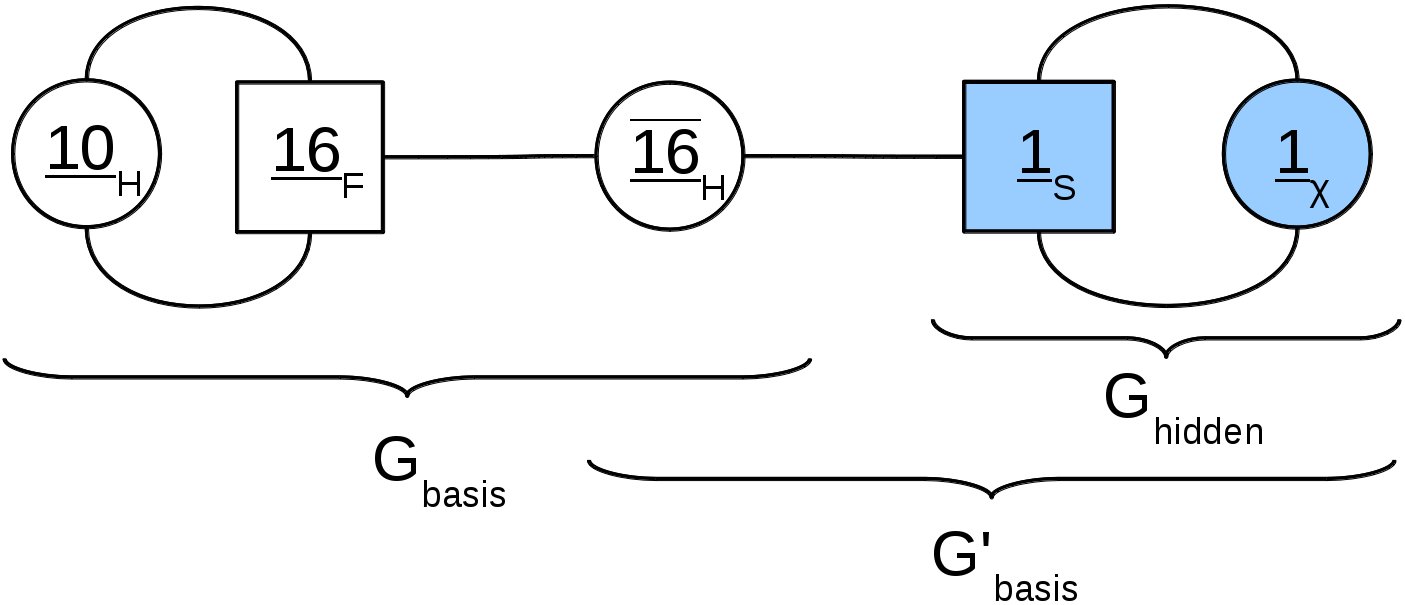}
\vspace*{10mm}\\
\includegraphics[width=0.6\textwidth]{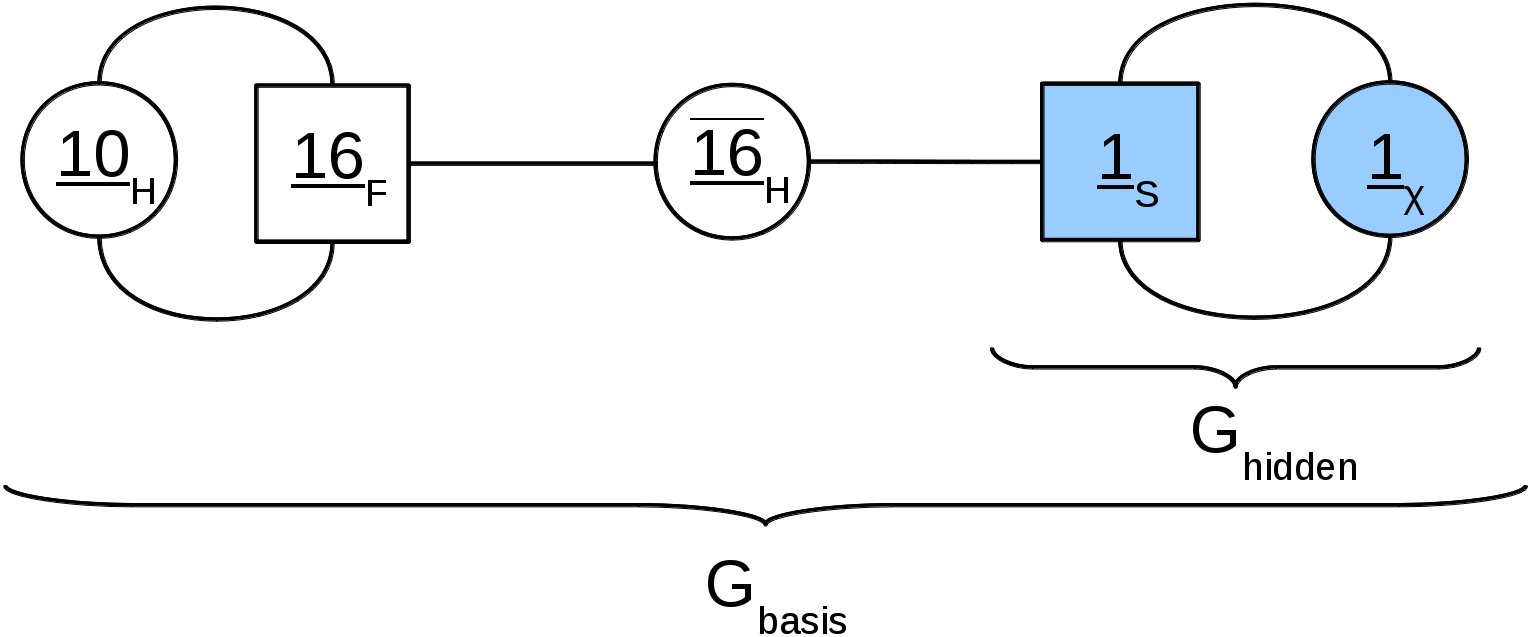}
\vspace*{10mm}\\
\includegraphics[width=0.6\textwidth]{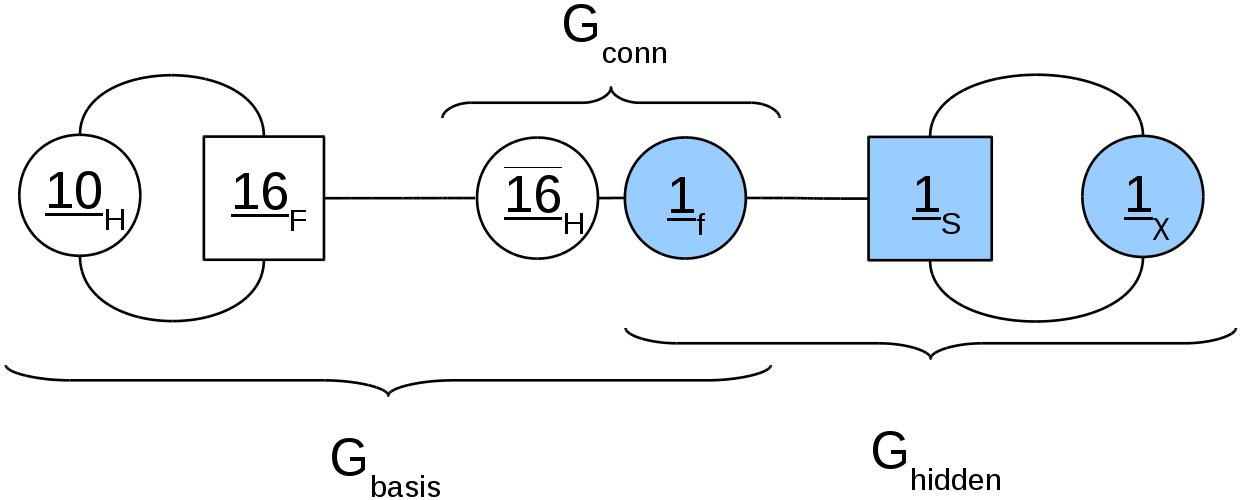}
\end{center}
\caption{Graphical representation of the Yukawa couplings and
the basis-fixing symmetry.
The visible sector interactions
are shown on the left-hand side of the diagrams by a circle connecting $\ten_H$
with $\sixt_F$. The portal interactions are symbolized by the lines connecting
$\one_S$ and $\sixt_F$ to $\overline{\sixt}_H$. The right-hand parts
of the diagrams show the hidden sector interactions between $\one_S$ and $\one_\chi$.
In the upper plot the basis information
is transferred to the hidden sector by $G_\text{basis}$ and $G_\text{basis}'$ with
the mediator fields being $\overline{\sixt}_H$. The figure in the middle shows the
direct communication of $G_\text{basis}$ via the singlets $\one_S$. The lower part
shows direct communication of $G_\text{basis}$ to the hidden sector via hidden sector
scalars $\one_f$.}\label{mediators}
\end{figure}

According to the structure of $\mathcal{L}^{(FS)}$ 
there are three basic possibilities for the choice of the mediator fields: 
\begin{itemize}
 \item
$\overline{\sixt}_H$ as mediators: For this  
at least three $\overline{\sixt}_H$ should be introduced that
have to transform under a symmetry $G_\text{basis}$
connecting it to the $\sixt_F$ and $G_\text{basis}'$
connecting it to the $\one_S$. The full basis-fixing symmetry is
$G_\text{basis} \times G_\text{basis}'$.
This case is illustrated in the upper part of Fig.~\ref{mediators}.
 \item
$\one_S$ as mediators: $\one_S$ have to transform under 
$G_\text{basis}$ on the top of $G_\text{hidden}$ (see the middle part of Fig.~\ref{mediators}).
In this case we have direct communication since $\one_S$ belong to the hidden sector.
 \item The role of the mediators can be played 
by flavons $\one_f$, which will allow to reduce the number of $\overline{\sixt}_H$, and 
open more flexibility for the structure of the hidden sector.
Namely, by means of an additional connecting symmetry $G_\text{conn}$, \textit{e.g.}\ $G_\text{conn} = \zed_m$,
one can ``bind'' flavons $\one_f$
to $\overline{\sixt}_H$ or $\one_S$, \textit{i.e.}\
\begin{equation}
\begin{split}
& \overline{\sixt}_{Hj}  \rightarrow \overline{\sixt}_H 
\left( \frac{\one_{fj}}{\Lambda} \right)^q 
\quad\quad\text{invariant under } G_\text{conn}\quad\text{or}\\
& \one_{Sj} \,  \rightarrow \one_S \left( \frac{\one_{fj}}{\Lambda} \right)^q
\quad\quad\text{invariant under } G_\text{conn}.
\end{split}
\label{bind_flavons}
\end{equation}
Here $j$ is the index which corresponds to the symmetry $G_\text{basis}$
and the power $q$ is a positive integer.
For example, if $\one_f$ transforms under $G_\text{basis}$ 
and $G_\text{hidden}$ and is connected to
$\overline{\sixt}_H$ via $G_\text{conn}$, 
the combination $\left( \overline{\sixt}_H \frac{\one_f}{\Lambda} \right)$
acts like a $\overline{\sixt}_H$-mediator.
Thus only one field $\overline{\sixt}_H$ is sufficient
and it does not need to obey any symmetries apart from $G_\text{conn}$ and SO(10).
Since the symmetry $G_\text{basis}$ directly acts on the hidden sector fields $\one_f$,
the communication of the basis information to the hidden sector is direct. The hidden
sector symmetry $G_\text{hidden}$ (under which the $\one_f$ have to be charged too) then
transmits the basis information also to the $\one_S$. This is illustrated
in the lower part of Fig.~\ref{mediators}.
\end{itemize}
%
%
%

\begin{figure}
\begin{center}
\includegraphics[width=0.7\textwidth]{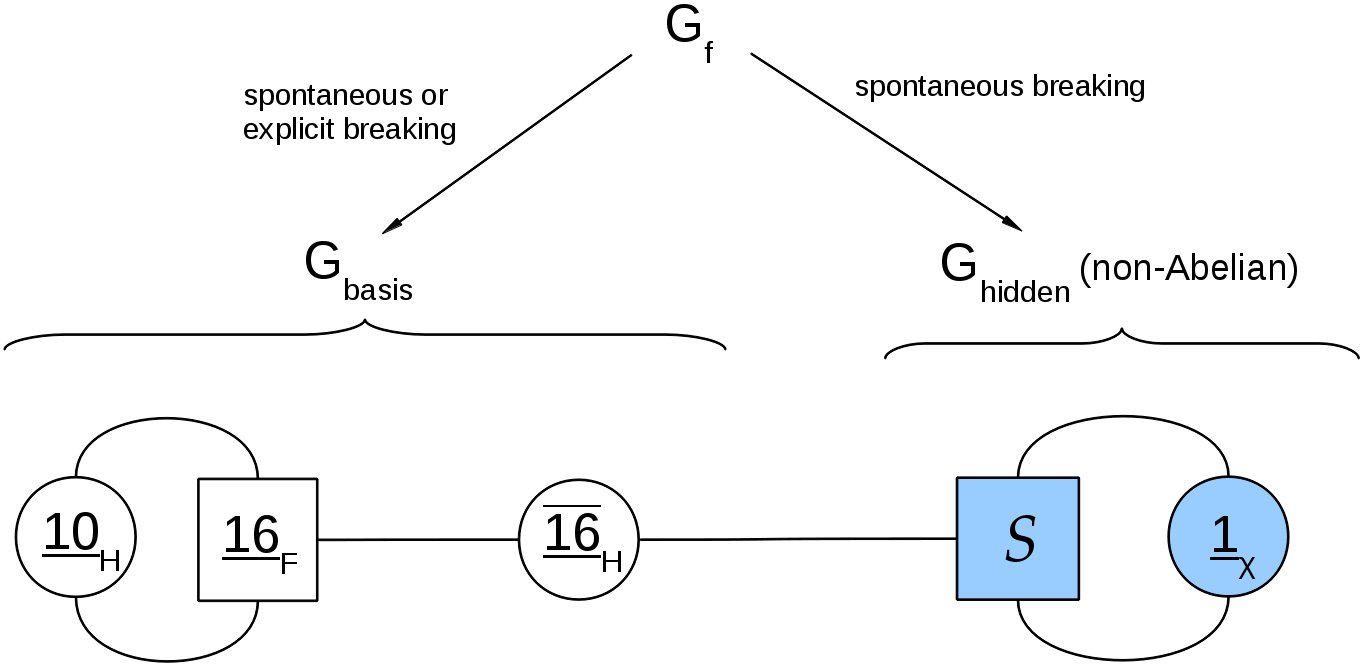}
\end{center}
\caption{$G_\text{basis}$ and $G_\text{hidden}$ as residual symmetries.}\label{Ghidden-nonAbelian}
\end{figure}

Let us now elaborate the required properties of $G_\text{hidden}$.
We differentiate two cases:
\begin{itemize}
 \item[A)] $G_\text{hidden}$ is an Abelian symmetry. In this case each individual
field $\one_S$ transforms as a one-dimensional representation of $G_\text{hidden}$
and the coupling to the $\sixt_F$ and $\overline{\sixt}_H$ in
the portal interaction $\mathcal{L}^{(FS)}$ can be invariant with
respect to all the symmetries.
 \item[B)] $G_\text{hidden}$ is a non-Abelian symmetry, in which case
some the fermionic singlets $\one_S$ form a multiplet $\mathcal{S}$ of an irreducible
representation of $G_\text{hidden}$. Then, all members of $\mathcal{S}$ have
to transform in the same way under all other symmetries, in particular also $G_\text{basis}^{(\prime)}$ which,
consequently, does not distinguish different members of $\mathcal{S}$.
Therefore, if $G_\text{hidden}$
is an exact symmetry of $\mathcal{L}^{(FS)}$,
communication of the basis-fixing symmetry to the hidden sector is excluded.
The only way to have a non-Abelian symmetry $G_\text{hidden}$ is thus
to break $G_\text{hidden}$ in the portal interaction $\mathcal{L}^{(FS)}$.
This breaking can be explicit or spontaneous (through additional flavons).
A scheme in which communication between the two sectors works can be conveniently arranged in the framework
of residual symmetries~\cite{lam1,lam2,grimus,toorop,hernandez,holthausen,fonseca}.
There, $G_\text{basis}$ and $G_\text{hidden}$ originate from a larger symmetry
$G_f$, which is broken to $G_\text{basis}$ in $\mathcal{L}^{(FF)}$ and $\mathcal{L}^{(FS)}$
and to $G_\text{hidden}$ in $\mathcal{L}^{(SS)}$---see Fig.~\ref{Ghidden-nonAbelian}.
Since now $G_\text{basis}$ and $G_\text{hidden}$
stem from the same larger symmetry group $G_f$,
even if $G_\text{hidden}$ is explicitly broken in $\mathcal{L}^{(FS)}$, the hidden sector
interaction $\mathcal{L}^{(SS)}$ ``knows'' about the basis fixed through
$G_\text{basis}$ and communication of flavour structures between the two sectors is
possible. From another point
of view: $G_\text{basis}$ and $G_\text{hidden}$ must be chosen in a way they can both
be embedded in a finite group $G_f$.
\end{itemize}

\section{Realizations of the framework}
\label{realizations}

In this section we present three realizations 
of the described framework. 
Note that we will not construct
full models, but essentially focus on the effects of the different symmetries.
The key elements are the same in all three cases:
\begin{enumerate}
 \item The symmetries $G_\text{basis}$ (and $G_\text{basis}'$) make $m_D$ and $M_{RS}$ diagonal 
(which selects $G_\text{basis} = \zed_2 \times \zed_2$), whereas $M_S$ is not diagonal.
The reason for this is
that in the portal interaction $\mathcal{L}^{(FS)}$ only two 
fields carry hidden sector (or basis-fixing symmetry) charges, whereas in $\mathcal{L}^{(SS)}$ there are three.
 \item We introduce an additional symmetry $G_\text{Yukawa}$ which is responsible
for the large hierarchy in $m_u \propto m_D$.\footnote{The
large hierarchy can be achieved in two ways. Either the hierarchy is
generated via flavon VEVs, or the couplings for the different 
generations are generated by operators of
different dimensions $\geq 4$. Also a combination of both mechanisms is possible.}
For this, the symmetry $G_\text{Yukawa}$ should distinguish three generations.
If the same symmetry also gives rise to the hierarchy of Yukawa-couplings in $\mathcal{L}^{(FS)}$,
cancellation between $m_D$ and $M_{RS}$ (complete or partial screening) is possible.
 \item The visible sector consists of three $\sixt_{Fa}$ $(a=1,2,3)$, 
two (or more) Higgs fields $\ten_H$ and one or three scalars $\overline{\sixt}_{H}$.
All $\ten_H$ have the same charges with respect to
$G_\text{basis}$.
Only one of them, $10_H^u$, gives
masses to the up-type quarks and neutrinos (see section~\ref{ckmphysics}).
 \item The hidden sector includes (among other components) three $\one_{Sj}$ $(j=1,2,3)$, 
one complex scalar $\one_{Y}$ responsible for the Yukawa-coupling hierarchy, and a set of complex scalars $\one_{\chi k}$.
 \item Since the Yukawa-coupling of the top-quark is $\mathcal{O}(1)$,
we produce the masses of the third generation of fermions at the renormalizable level.
The couplings for the first and second generation will be produced through effective
operators of higher dimension.
Therefore, the symmetries
$G_\text{basis}$ and $G_\text{Yukawa}$ must be Abelian, or, if non-Abelian, act
on the third generation with a one-dimensional representation.\\
For models with three $\sixt_F$ transforming under a three-dimensional irreducible
representation of a discrete group, and which realize screening,
see \textit{e.g.}~\cite{Hagedorn}.
\end{enumerate}

\subsection{Realization I: $\one_S$ mediators  with Abelian hidden symmetries}
\label{ill1}

\begin{figure}
\begin{center}
\includegraphics[width=0.75\textwidth]{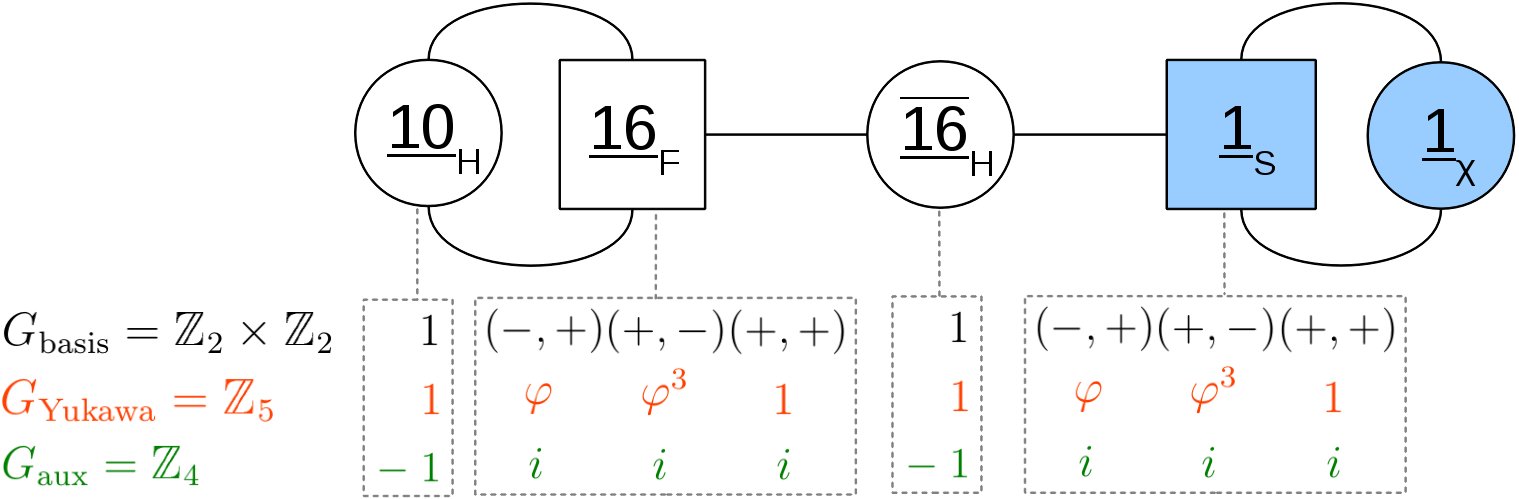}
\end{center}
\caption{Graphical representation of the charge assignments for realization~I
($\varphi \equiv e^{2\pi i/5}$). The scalar $\one_Y$ is not shown here. It transforms
as $\one_Y \rightarrow \varphi^4\,\one_Y$ under $G_\text{Yukawa}$ and
trivially under $G_\text{basis}$ and $G_\text{aux}$.
The charges of the flavon fields $\one_\chi$ are shown in table~\ref{flavons}.}\label{fig-illustrations12}
\end{figure}

The fields communicating the basis-fixing
symmetry to the hidden sector are chosen to be the heavy singlets
$\one_S$ themselves.
The charge assignments under the symmetries are shown
in Fig.~\ref{fig-illustrations12}.
%
%
%
The fact that all fermions transform in
exactly the same way under the non-gauged symmetries
makes the present scenario particularly appealing. This could be a remnant of further unification
beyond SO(10)~\cite{screening2}.
If $\sixt_F$ and $\one_S$ stem from the decomposition 
of an $\mathrm{E}_6$-multiplet~\cite{E6-1,E6-2,E6-3,E6-4} into SO(10)-multiplets 
\begin{equation}
\underline{27} \rightarrow \one  \oplus \ten
\oplus \sixt,
\end{equation}
we automatically obtain the same number of 16-plet and singlet fermions
having the same transformation properties under all discrete groups.\footnote{In
order not to generate unwanted new interactions, the ten-plet fermions should 
be either at Planck scale or not realized in the
theory at all.} Moreover, since 
$\mathcal{L}^{(FF)}$ and $\mathcal{L}^{(FS)}$ originate from the same coupling
in the underlying $\mathrm{E}_6$-theory,\footnote{This requires that also the scalars $\ten_H$ and
$\overline{\sixt}_H$ are embedded in an $\mathrm{E}_6$-multiplet via\\
$\overline{\underline{27}} \rightarrow \one \oplus \ten
\oplus \overline{\sixt}$.}
also the Yukawa-coupling constants are the same, \textit{i.e.}\ $Y^{(FF)}_u=Y^{(FS)}$.

The symmetry $G_\text{basis}$ makes $Y^{(FF)}_u$ and $Y^{(FS)}$
diagonal. The strong mass hierarchy of quarks and charged leptons is achieved via effective operators
of different dimensions for the different generations. The minimal group which 
can provide the required hierarchy is 
$G_\text{Yukawa} = \zed_5$. The charges with respect to this group 
determine the dimension of the effective operator.
Thus, taking into account operators up to dimension six, we obtain
\begin{equation}
\label{Yukawa-hierarchy}
Y^{(FF)}_u = Y^{(FS)} = \mathrm{diag}\left( 
y_1 \left(\frac{\langle\one_Y\rangle}{\Lambda}\right)^2,\enspace
y_2 \frac{\langle\one_Y\rangle}{\Lambda},\enspace
y_3 \right). 
\end{equation}
Here we have neglected the dimension-six contribution
$\tilde{y}_3 \frac{\langle\one_Y\rangle \langle\one_Y^\ast\rangle}{\Lambda^2}$
to the third-generation Yukawa coupling.
The mass matrices $m_D$ and $M_{RS}$ are strictly proportional to each other and there is
exact screening
\begin{equation}
D = m_D (M_{RS}^{-1})^T \propto \bone_3.
\end{equation}
The neutrino mass matrix is then given by
\begin{equation}
m_\nu^{DS} = D M_{S} D^T \propto M_{S}.
\end{equation}
The auxiliary symmetry $G_\text{aux}$ forbids a bare mass term for the singlets $\one_S$
and the dimension-5 contribution
$\frac{1}{\Lambda}\sixt_F \sixt_F \overline{\sixt}_{H} \overline{\sixt}_H$
to the right-handed neutrino mass term $M_R$ in the mass matrix~(\ref{massmat}). 
Introduction of flavons transforming as shown in table~\ref{flavons}
allows to generate independently any element of $M_{S}$,
and therefore to obtain any set of texture zeros in $M_{S}$.
For example, introduction of $\one_{\chi 22}$, $\one_{\chi 23}$ and
$\one_{\chi 33}$ only, will produce a dominant 2-3-block. Notice that
$G_\text{Yukawa}$ plays an important role in structuring $M_S$.
\begin{table}
\begin{center}
\begin{tabular}{|c|c|c|c|}
\hline
Flavon & $\zed_2\times \zed_2$ & $\zed_5$ & $\zed_4^\text{aux}$\\
\hline
$\one_{\chi 11}$ & $(+,+)$ & $\varphi^3$ & $-1$ \\
$\one_{\chi 12}$ & $(-,-)$ & $\varphi$ & $-1$ \\
$\one_{\chi 13}$ & $(-,+)$ & $\varphi^4$ & $-1$ \\
$\one_{\chi 22}$ & $(+,+)$ & $\varphi^4$ & $-1$ \\
$\one_{\chi 23}$ & $(+,-)$ & $\varphi^2$ & $-1$ \\
$\one_{\chi 33}$ & $(+,+)$ & $1$ & $-1$ \\
\hline
\end{tabular}
\caption{Transformation properties of the flavons generating $M_{S}$.}\label{flavons}
\end{center}
\end{table}
In order to get $\theta_{13}^X = 0$ without fine-tuning, a non-Abelian structure in $M_S$
is needed (see section~\ref{ill3}). 

With more flavons an additional Abelian symmetry $G_\text{hidden}$
can be realized, \textit{e.g.}\ $G_\text{hidden} = \zed_n$. If the singlets
$\one_{Si}$ ($i=1,2,3$) transform with charges $\gamma_i$ and flavons $\one_{fi}$
transform with charges $n-\gamma_i$, in $\mathcal{L}^{(FS)}$ the singlets $\one_{Si}$ can be substituted
by operators
\begin{equation}
\one_{Si}\left(\frac{1_{fi}}{\Lambda}\right)
\end{equation}
without breaking of any other symmetry---see also Eq.~(\ref{bind_flavons}).
Then, even if all flavons of table~\ref{flavons} are present in the theory,
the extended hidden symmetry could for example be used to obtain a dominant
2-3-block in $M_S$, \textit{i.e.}\
\begin{equation}
M_S \sim
\begin{pmatrix}
a & 0 & 0\\
0 & b & d\\
0 & d & c
\end{pmatrix}
\end{equation}
giving large 2-3-mixing in $U_X$.
For this already $G_\text{hidden}=\zed_2$ with charge assignment $\one_S \sim (-,+,+)$
would be sufficient. Then nonzero 12 and 13 elements
of $M_S$ can be generated by additional flavons or interactions with
other hidden sector fermions (see section~\ref{additional_singlets}),
or by higher order operators.

\subsection{Realization II:  $\one_S$ mediators with broken non-Abelian symmetry $G_f$}
\label{ill2}

The field content and the symmetries 
$G_\text{basis}$ and $G_\text{Yukawa}$ 
are the same as in realization~I.
In addition, now we introduce a non-Abelian symmetry $G_\text{hidden}$
in the basis fixed by $G_\text{basis}$.
For this we embed $G_\text{basis}$ and $G_\text{hidden}$ into an extended flavour symmetry group
$G_f \supset G_\text{basis},\,G_\text{hidden}$.
The embedding into the same group $G_f$ ensures that
$G_\text{hidden}$ is introduced in the basis fixed by $G_\text{basis}$. This is necessary
to communicate information about $G_\text{hidden}$ to the visible sector.
The flavons $\one_\chi$ break $G_f$
spontaneously\footnote{Since the breaking of $G_f$ in $\mathcal{L}^{(SS)}$
happens at a very high energy scale of $\sim10^{18}\,\text{GeV}$, we want this breaking to be spontaneous through $\one_\chi$
rather than explicit.} to $G_\text{hidden}$ in the hidden sector
interaction $\mathcal{L}^{(SS)}$.
In $\mathcal{L}^{(FS)}$ (as well as in $\mathcal{L}^{(FF)}$) $G_f$ is broken down to $G_\text{basis}$
explicitly or spontaneously (the latter would require a substantial
complication of the model).
In this way also $G_\text{hidden}$
is broken explicitly in the low-energy interactions 
(see Fig.~\ref{Ghidden-nonAbelian}).
Notice that $G_\text{basis}$ is unbroken in $\mathcal{L}^{(FF)}$
and $\mathcal{L}^{(FS)}$. It will be broken by the mechanism which
generates the CKM-mixing---see section~\ref{ckmphysics}.

In what follows, we consider the 2-3-permutation symmetry ($\mu\tau$-symmetry) as $G_\text{hidden}$.
The minimal group which realizes the embedding
is\footnote{By $\mathrm{D}_4$ we denote the
dihedral group~\cite{dihedral} of order eight. 
Sometimes in the literature this group is also denoted by $\mathrm{D}_8$.} $G_f = \mathrm{D}_4\times \zed_2$
with three generators $A$, $B$ and $C$
and the faithful three-dimensional reducible representation
\begin{equation}\label{generators}
\mathbf{\underline{3}}:\quad A\mapsto
\begin{pmatrix}
-1 & 0 & 0\\
 0 & 1 & 0\\
 0 & 0 & 1
\end{pmatrix},\quad
B\mapsto
\begin{pmatrix}
 1 & 0 & 0\\
 0 &-1 & 0\\
 0 & 0 & 1
\end{pmatrix},\quad
C \mapsto
\begin{pmatrix}
 1 & 0 & 0\\
 0 & 0 & 1\\
 0 & 1 & 0
\end{pmatrix}.
\end{equation}
We assign $\sixt_F$ and $\one_S$ to transform under reducible triplet representations $\mathbf{\underline{3}}$
of $G_f$. $A$ and $B$ alone generate $G_\text{basis} = \zed_2\times\zed_2$,
so that explicit breaking $G_f \rightarrow G_\text{basis}$
leads to diagonal $Y^{(FF)}_u$ just as in the previous section.
Equality $Y^{(FF)}_u=Y^{(FS)}$ can again be achieved by embedding of SO(10) into
$\mathrm{E}_6$.
The generator $C$ corresponds to the 2-3-permutation symmetry. 

The irreducible representations of $\mathrm{D}_4\times \zed_2$ are
\begin{subequations}
\begin{align}
 & \mathbf{\one_1}: \quad A\mapsto 1,\quad B\mapsto 1,\quad C\mapsto 1,\\
 & \mathbf{\one_2}: \quad A\mapsto 1,\quad B\mapsto -1,\quad C\mapsto 1,\\
 & \mathbf{\one_3}: \quad A\mapsto 1,\quad B\mapsto 1,\quad C\mapsto -1,\\
 & \mathbf{\one_4}: \quad A\mapsto 1,\quad B\mapsto -1,\quad C\mapsto -1,\\
 & \mathbf{\one_1'}: \quad A\mapsto -1,\quad B\mapsto 1,\quad C\mapsto 1,\\
 & \mathbf{\underline{2}}: \quad
A\mapsto
\begin{pmatrix}
 1 & 0\\
 0 & 1
\end{pmatrix},\quad
B\mapsto
\begin{pmatrix}
 -1 & 0\\
  0 & 1
\end{pmatrix},\quad
C \mapsto
\begin{pmatrix}
 0 & 1\\
 1 & 0
\end{pmatrix}
\end{align}
\end{subequations}
and the products $\mathbf{\underline{2}'} \equiv \mathbf{\one_1'} \otimes \mathbf{\underline{2}}$
and $\mathbf{\one_i'} \equiv \mathbf{\one_1'} \otimes \mathbf{\one_i}$ $(i=2, 3, 4)$.
The relevant tensor product for the construction of the hidden sector interaction $\mathcal{L}^{(SS)}$
is thus
\begin{equation}
\mathbf{\underline{3}} \otimes \mathbf{\underline{3}} =
(\mathbf{\one_1'} \oplus \mathbf{\underline{2}}) \otimes (\mathbf{\one_1'} \oplus \mathbf{\underline{2}}) =
\underbrace{\mathbf{\one_1'} \otimes \mathbf{\one_1'}}_{\mathbf{\one_1}}
\oplus \underbrace{\mathbf{\one_1'} \otimes \mathbf{\underline{2}}}_{\mathbf{\underline{2}'}}
\oplus \underbrace{\mathbf{\underline{2}} \otimes \mathbf{\one_1'}}_{\mathbf{\underline{2}'}} \oplus
\underbrace{\mathbf{\underline{2}} \otimes \mathbf{\underline{2}}}_{
 \mathbf{\one_1} \oplus
 \mathbf{\one_2} \oplus
 \mathbf{\one_3} \oplus
 \mathbf{\one_4}}.
\end{equation}
Introducing singlet flavons $\chi\sim \mathbf{\one_1}$, $\rho\sim \mathbf{\one_2}$ and a
flavon doublet $\eta = (\eta_1,\eta_2)^T \sim \mathbf{\underline{2}'}$, we obtain the Yukawa-interaction\footnote{Introduction
of a singlet $\mathbf{\one_3}$ would lead to non-equality of the 22 and 33 elements of $M_S$ and
$\mathbf{\one_4}$ gives an antisymmetric contribution to $M_S$ which vanishes due to the Majorana nature
of $\one_S$.} invariant with respect to $G_f$
\begin{equation}
\mathcal{L}^{(SS)} = -\frac{1}{2}
\begin{pmatrix} \one_{S1} & \one_{S2} & \one_{S3} \end{pmatrix}
\begin{pmatrix}
 y_\chi \chi & y_\eta \eta_1 & y_\eta \eta_2\\
 y_\eta \eta_1 & y_\chi' \chi & y_\rho \rho\\
 y_\eta \eta_2 & y_\rho \rho  & y_\chi' \chi
\end{pmatrix}
\begin{pmatrix} \one_{S1} \\ \one_{S2} \\ \one_{S3} \end{pmatrix} + \hc
\end{equation}
So, if $\langle \chi \rangle \neq 0$, $\langle \rho \rangle \neq 0$
and the doublet VEVs are aligned as $\langle \eta_1 \rangle = \langle \eta_2 \rangle$,
$m_\nu^{DS}$ will be the most general 2-3-permutation symmetric matrix
\begin{equation}\label{MS-D4}
m_\nu^{DS} \propto M_S =
\begin{pmatrix}
a & b & b\\
b & c & d\\
b & d & c
\end{pmatrix},
\end{equation}
which is compatible with the
neutrino mass squared-differences for both mass orderings
and gives $\theta_{23}^X = 45^\circ$ and $\theta_{13}^X = 0^\circ$. Also the required
size of 1-2-mixing can be obtained.
Since in this paper we focus on the symmetries and do not construct models,
we do not discuss mechanisms to get the required vacuum alignment. Such
mechanisms have been elaborated in the literature, see \textit{e.g.}~\cite{Altarelli-Feruglio,Holthausen},
and can be realized here.

Also the group $G_\text{Yukawa} = \zed_5$ can be embedded into a discrete group
$G_f \supset G_\text{basis} \times G_\text{Yukawa}$ by adding
a fourth generator
\begin{equation}
D\mapsto
\begin{pmatrix}
 \varphi & 0 & 0\\
 0 & \varphi^3 & 0\\
 0 & 0 & 1
\end{pmatrix}
\end{equation}
to Eq.~(\ref{generators}).
Using the computer algebra system GAP~\cite{GAP} we find that the resulting
group has the structure $G_f = \zed_{10} \times (\zed_{10} \times \zed_2) \rtimes \zed_{2}$. By construction, it contains
$G_\text{basis} \times G_\text{Yukawa} = \zed_2\times\zed_2\times\zed_5$ as a subgroup.
Since the three-dimensional representation of the extended group $G_f$ is still reducible,
\textit{i.e.}\ $\mathbf{\underline{3}} = \mathbf{\one'} \oplus \mathbf{\underline{2}}$,
generation of the 12 and 13 elements of $M_S$ needs a scalar doublet $\eta\sim (\mathbf{\one'}\otimes \mathbf{\underline{2}})^\ast$.
The 11-element can be generated by a coupling
with a flavon $\chi\sim (\mathbf{\one'}\otimes \mathbf{\one'})^\ast$. Finally, the 2-3-block of $M_S$ is
determined by the tensor product
\begin{equation}
\mathbf{\underline{2}} \otimes \mathbf{\underline{2}} = \mathbf{\one_s} \oplus \mathbf{\one_a} \oplus \widetilde{\mathbf{\underline{2}}},
\end{equation}
where
\begin{subequations}
\begin{align}
 & \mathbf{\one_s}: \quad A\mapsto 1,\quad B\mapsto -1,\quad C\mapsto  1,\quad D\mapsto \varphi^3,\\
 & \mathbf{\one_a}: \quad A\mapsto 1,\quad B\mapsto -1,\quad C\mapsto -1,\quad D\mapsto \varphi^3,\\
 & \widetilde{\mathbf{\underline{2}}}: \quad
A\mapsto
\begin{pmatrix}
 1 & 0\\
 0 & 1
\end{pmatrix},\quad
B\mapsto
\begin{pmatrix}
  1 & 0\\
  0 & 1
\end{pmatrix},\quad
C \mapsto
\begin{pmatrix}
 0 & 1\\
 1 & 0
\end{pmatrix},\quad
D \mapsto
\begin{pmatrix}
 \varphi & 0\\
 0 & 1
\end{pmatrix}.
\end{align}
\end{subequations}
The representation $\mathbf{\one_a}$ is the antisymmetric component of $\mathbf{\underline{2}} \otimes \mathbf{\underline{2}}$
and therefore does not contribute to $\mathcal{L}^{(SS)}$. The off-diagonal elements of the 2-3-block of $M_S$
can be generated via the Yukawa interaction with a singlet $\rho\sim \mathbf{\one_s}^\ast$ and the 22 and 33 elements
need a flavon doublet $\chi' = (\chi_1',\chi_2')^T\sim \widetilde{\mathbf{\underline{2}}}^\ast$. In this case, the hidden sector
interactions obtain the form
\begin{equation}
\mathcal{L}^{(SS)} = -\frac{1}{2}
\begin{pmatrix} \one_{S1} & \one_{S2} & \one_{S3} \end{pmatrix}
\begin{pmatrix}
 y_\chi \chi & y_\eta \eta_1 & y_\eta \eta_2\\
 y_\eta \eta_1 & y_\chi' \chi_1' & y_\rho \rho\\
 y_\eta \eta_2 & y_\rho \rho  & y_\chi' \chi_2'
\end{pmatrix}
\begin{pmatrix} \one_{S1} \\ \one_{S2} \\ \one_{S3} \end{pmatrix} + \hc
\end{equation}
If both doublet VEVs break $G_f$ to 2-3-permutation symmetry, \textit{i.e.}\ $\langle \eta_1\rangle = \langle \eta_2\rangle$
and $\langle \chi_1'\rangle = \langle \chi_2'\rangle$, $M_S$ will be 2-3-permutation symmetric.

Instead of amending $\zed_2\times\zed_2$ directly by the 2-3-permutation symmetry, we could also
use the permutation symmetry
\begin{equation}
C \mapsto
\begin{pmatrix}
 0 & 1 & 0\\
 0 & 0 & 1\\
 1 & 0 & 0
\end{pmatrix}
\end{equation}
which yields $G_f = \mathrm{A}_4 \times \zed_2$, \textit{i.e.}\ a group with a three-dimensional irreducible representation.
Embedding also $G_\text{Yukawa} = \zed_5$
extends the flavour group to $G_f = \Delta(3\times 10^2) \times \zed_{10}$, \textit{i.e.}\ to a direct product
of a cyclic group and a dihedral-like~\cite{Bovier,Luhn} subgroup of SU(3). In both cases the 2-3-permutation symmetry
can be achieved by alignment of flavon triplet VEVs of $G_f$ similar to the previous case.

Let us finally consider breaking of $G_f$ in $\mathcal{L}^{(FS)}$ which implies
the explicit breaking of $G_\text{hidden}$. This breaking will affect $M_S$ and
consequently $m_\nu^{DS}$.
We can estimate the corrections to $M_S$ and $m_\nu$
using the general results of the type-I seesaw expansion~\cite{Grimus-Lavoura}.
According to~\cite{Grimus-Lavoura} we expect that
the effect of explicit breaking of $G_\text{hidden}$ on $M_S$ is of the
order of magnitude of $\delta M_S \sim M_\text{GUT}^2/m_S \sim 10^{-4} m_S$.
Consequently, in the double seesaw
expression, the effect will be of the order of $\delta m_\nu^{DS}/m_\nu^{DS} \sim 10^{-4}$, \textit{i.e.}\
the corrections are negligible.

\subsection{Realization III: Scalar fields $\overline{\sixt}_{H}$ as mediators}
\label{ill3}

\begin{figure}
\begin{center}
\includegraphics[width=0.90\textwidth]{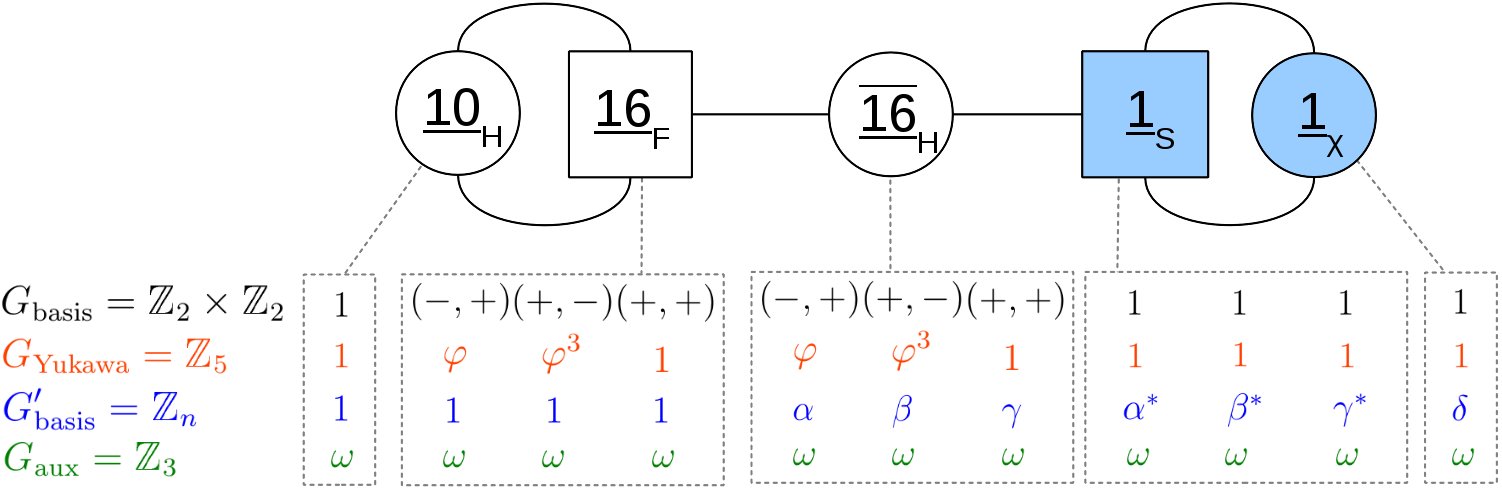}
\end{center}
\caption{Graphical representation of the couplings and charge assignments for realization~III.
The numbers $\alpha$, $\beta$, $\gamma$ and $\delta$ are $n$-th roots of unity.
($\varphi = e^{2\pi i/5}$, $\omega\equiv e^{2\pi i/3}$).}
\label{fig-illustration3}
\end{figure}

In this case we need to introduce three $\overline{\sixt}_{H}$. 
The symmetries and charge assignments are shown in Fig.~\ref{fig-illustration3}.
%
%
%
%
The $\overline{\sixt}_H$
have the same transformation properties under $G_\text{basis}$
and $G_\text{Yukawa}$ as $\sixt_F$.  
The symmetry $G_\text{basis}$ makes the couplings $\sixt_F \sixt_F$
and $\sixt_F\overline{\sixt}_H$ diagonal and 
$G_\text{Yukawa}$ generates the strong hierarchy in
$m_D$ and $M_{RS}$. Thus, $\mathcal{L}^{(FF)}$ is given by~(\ref{Yuk1}) with
the Yukawa coupling $Y^{(FF)}_u$ of~(\ref{Yukawa-hierarchy}).
Note, however, that since embedding into $\mathrm{E}_6$ is not possible here,
the Yukawa couplings $Y^{(FF)}_u$ and $Y^{(FS)}$ are not equal and screening is
in general only partial.
$D$ is still diagonal with, however, non-equal elements. This can be used
to explain some features of mixing.  

Notice that now communication between the visible and the hidden sector is not
direct---it proceeds via $\overline{\sixt}_H$. Furthermore, just $G_\text{basis}=\zed_2\times\zed_2$
is not enough since all hidden sector fields are singlets of $G_\text{basis}$.
To fix a basis in both sectors, an additional symmetry $G_\text{basis}'$ is required under
which both $\overline{\sixt}_H$ and hidden sector fields $\one_S$ are transformed.
So, the information about the basis is transferred in two steps: from $\sixt_F$ to $\overline{\sixt}_H$
by $G_\text{basis}$ and from $\overline{\sixt}_H$ to $\one_S$ by $G_\text{basis}'$. $\overline{\sixt}_H$
is charged with respect to both $G_\text{basis}$ and $G_\text{basis}'$.

As $G_\text{basis}'$ we use an Abelian symmetry.
For simplicity we choose $G_\text{basis}'=\zed_n$ but the following arguments
hold for any Abelian group.
If the $\zed_n$ charges of 
$\one_S$ are  $\alpha\neq \beta \neq \gamma \neq \alpha$, it makes also the
couplings $\overline{\sixt}_{H}\one_S$ diagonal---each $\overline{\sixt}_{H}$
is uniquely connected to one $\one_S$. Therefore, due to the
mediation by $\overline{\sixt}_H$, 
the $\one_S$ ``know'' about the basis choice in the space of $\sixt_F$. For
$\mathcal{L}^{(FS)}$ we obtain
\begin{equation}\label{LFS-final}
 \mathcal{L}^{(FS)} = - \left[
 y_1' \left(\frac{\one_Y}{\Lambda}\right)^2 \sixt_{F1} \one_{S1} \overline{\sixt}_{H1} +
 y_2' \frac{\one_Y}{\Lambda} \sixt_{F2} \one_{S2} \overline{\sixt}_{H2} +
 y_3' \sixt_{F3} \one_{S3} \overline{\sixt}_{H3} \right] + \hc,
\end{equation}
and the matrix $D = m_D (M_{RS}^{-1})^T$ is given by
\begin{equation}
D = \mathrm{diag}\left(
\frac{y_1 v_{10}^u}{y_1' v_{16,1}},\,
\frac{y_2 v_{10}^u}{y_2' v_{16,2}},\,
\frac{y_3 v_{10}^u}{y_3' v_{16,3}}
\right).
\end{equation}
The simplest way to obtain partial screening is to assume that all Yukawa couplings 
are of similar size and that the VEVs of the three $\overline{\sixt}_H$ are of the same
size: $v_{16,1} \sim v_{16,2} \sim v_{16,3}$.
In this case $D$ is a diagonal matrix with elements of the same order.
The dimension-three bare mass term for the $\one_S$
and dimension-six couplings
of the form $\frac{1}{\Lambda}\sixt_F \sixt_F \overline{\sixt}_{H} \overline{\sixt}_H$, which would give rise
to a right-handed neutrino mass term $M_R \neq 0$ in Eq.~(\ref{massmat}),
are forbidden due to the auxiliary symmetry $G_\text{aux} = \zed_3$.\footnote{We could also have used
$G_\text{aux} = \zed_4$ as in realizations~I and~II. The main difference between these two choices is
in the scalar potential, where $\zed_4$ forbids cubic scalar couplings, while $\zed_3$ does not.}

As discussed in section~\ref{section-mediators}, the role of $\overline{\sixt}_{Hi}$ as the mediators
could be given to new flavons $\one_{fi}$ having the transformation properties
of $\overline{\sixt}_{Hi}$. Then only one scalar 16-plet
is needed which should 
transform trivially under all discrete groups except for a connecting symmetry
$G_\text{conn}$. In the present example we can for instance choose
\begin{equation}
G_\text{conn}:\quad \one_f \rightarrow \omega \one_f,\quad
\overline{\sixt}_H \rightarrow \omega^2 \overline{\sixt}_H.
\end{equation}
Then in Eq.~(\ref{LFS-final}) $\overline{\sixt}_{Hi}$ should 
be replaced by $\overline{\sixt}_{H}\left(\frac{\one_{fi}}{\Lambda}\right)$.
In order to achieve partial screening we should require
$\langle\one_{f1}\rangle\sim\langle\one_{f2}\rangle\sim\langle\one_{f3}\rangle$.

As in realization~I, by choosing appropriate $\zed_n$ charges $\delta_i$ for the
flavons $\one_{\chi i}$, we can
generate texture zeros in $M_S$. Since  $G_\text{hidden} = \zed_n$ 
can be extended to an arbitrary Abelian symmetry,
all types of texture zeros in $M_S$ can be obtained~\cite{Grimus-Joshipura}.
By means of texture zeros the exact 2-3-permutation symmetry in $M_S$ can only be achieved for
the matrix
\begin{equation}
M_S =
\begin{pmatrix}
 a & 0 & 0\\
 0 & 0 & b\\
 0 & b & 0
\end{pmatrix}.
\end{equation}
Then for the light neutrinos we obtain
\begin{equation}
m_\nu^{DS} =
\begin{pmatrix}
 a D_{11}^2 & 0 & 0\\
 0 & 0 & b D_{22} D_{33}\\
 0 & b D_{22} D_{33} & 0
\end{pmatrix}.
\end{equation}
This matrix can be experimentally feasible for a quasi-degenerate neutrino mass spectrum
only, \textit{i.e.}\ when $|a D_{11}^2| \approx |b D_{22} D_{33}|$, and
in addition corrections
to $M_S$ are needed to generate 1-2-mixing and mass splitting.

An approximate 2-3-permutation symmetric 
mass matrix $M_S$ may for example be
\begin{equation}
M_{S} =
\begin{pmatrix}
 a & c & d\\
 c & 0 & b\\
 d & b & 0
\end{pmatrix}.
\end{equation}
If $d=c(1+\epsilon)$, the eigenvector $(0,-1,1)^T$ of
$m_\nu m_\nu^\dagger$ for an exactly 2-3-permutation symmetric $m_\nu$ will get corrections of order $\epsilon$
in all three entries. Consequently, $\mathrm{sin}\,\theta_{13}^X \sim |(c-d)/c|$, \textit{i.e.}\
for $\mathrm{sin}\,\theta_{13}^X \ll \lambda \approx 0.2$ a fine tuning at the few-percent-level
is  necessary. Moreover, also the screening matrix $D$  has to be close  to
$\bone_3$ at the percent level to maintain the approximate 
2-3-permutation symmetry in $m_\nu^{DS}$.  

To summarize, the scenario with a purely Abelian hidden sector symmetry, as expected,
needs fine-tuning at the percent level to obtain the relation~(\ref{quark-lepton-mixing}).
Non-Abelian structures in $M_S$ can
be introduced with explicit symmetry
breaking, as in realization~II or, possibly, through the effects 
of additional SO(10)-singlet fermions,
which are discussed in section~\ref{additional_singlets}.

\section{Effects of additional SO(10)-singlet fermions}
\label{additional_singlets}

Up to now we have considered three SO(10)-singlet fermions
$S$ which couple directly to the $\sixt_F$. In this section we will discuss
the effects of additional fermionic singlets from the hidden sector.
We differentiate two basic cases:
\begin{enumerate}
 \item Additional singlets which directly couple to the $\sixt_F$.
 \item Additional singlets which do not directly couple to the $\sixt_F$ but mix
with the other fermionic singlets of the hidden sector.
\end{enumerate}
In the first case, the number of singlets $S$ directly coupled to the
$\sixt_F$ in the portal interaction $\mathcal{L}^{(FS)}$ is larger than three.
Thus, $M_{RS}$ is not a square matrix and therefore not invertible.
However, if $M_S$ is invertible, $M_R$ is given by the usual expression
$M_R \approx - M_{RS} M_S^{-1} M_{RS}^T$ and the seesaw formulae change to
\begin{subequations}\label{general-seesaw}
\begin{align}
 & m_\nu \approx m_\nu^{DS} + m_{\nu}^{LS} + m_\nu' \\
 & m_\nu^{DS} = -m_D M_R^{-1} m_D^T, \\
 & m_{\nu}^{LS} = m_D M_R^{-1} M_{RS} M_S^{-1} m_{\nu S}^T + 
m_{\nu S} M_S^{-1} M_{RS}^T M_{R}^{-1} m_D^T, \\
 & m_\nu' = -m_{\nu S} M_S^{-1} \left\{
\bone + M_{RS}^T M_R^{-1} M_{RS} M_S^{-1}
\right\} m_{\nu S}^T.
\end{align}
\end{subequations}
With the mass scales indicated in table~\ref{scales},
the new term $m_\nu'$
is expected to be of the order $\sim 10^{-5}\,\text{eV}$ and therefore negligible
compared to all other contributions to $m_\nu$.
If $M_S$ is singular and $M_{RS}$ has rank three, the matrix
\begin{equation}
\begin{pmatrix}
 0 & M_{RS}\\
 M_{RS}^T & M_{S}
\end{pmatrix}
\end{equation}
can still be invertible, but the formulae~(\ref{MReff})-(\ref{LS})
and~(\ref{general-seesaw}) will not hold any more.
Notice that the case of more than three singlets coupled directly
to the $\sixt_F$ is disfavoured by the requirement of screening
of the Dirac mass matrix.
Indeed, in the basis where $m_D$ is diagonal, exact screening
in the sense that the structure of $m_\nu^{DS}$ is solely determined by
$M_S$ requires
\begin{equation}
M_{RS} \propto
\begin{pmatrix}
m_u & 0 & 0 & 0 & \ldots & 0 \\
0 & m_c & 0 & 0 & \ldots & 0 \\
0 & 0 & m_t & 0 & \ldots & 0
\end{pmatrix},
\end{equation}
\textit{i.e.}\ a diagonal $M_{RS}$.
However, if $M_{RS}$ is a general matrix with hierarchy among
its rows, partial screening is possible.

For case~2, which is favoured by screening, we need
vanishing of the couplings of the additional singlets $\one_S'$ to the $\sixt_F$.
This can be achieved for example by introduction of an additional $\zed_2$ symmetry
under which $\sixt_F$ and the three singlets $\one_S$ change the sign 
whereas $\one_S'$ do not change. In order to allow couplings between $\one_S$ and $\one_S'$
also new flavon fields charged under $\zed_2$ should be introduced.
The $\zed_2$-symmetry thus acts as a kind of connecting symmetry. Among all
hidden sector fields it selects three which can directly couple to the $\sixt_F$.
In this case the neutrino mass
matrix has the form 
\begin{equation}
\mathcal{M} =
\begin{pmatrix}
0 & m_D & m_{\nu S} & 0\\
m_D^T & 0 & M_{RS} & 0\\
m_{\nu S}^T & M_{RS}^T & A & B\\
0 & 0 & B^T & C
\end{pmatrix}
\end{equation}
and the double seesaw formula becomes
\begin{equation}\label{modified-DS}
 m_\nu^{DS} = m_D M_{RS}^{-1 T}  \left( A - BC^{-1}B^T \right)  M_{RS}^{-1} m_D^T,
\end{equation}
while the linear seesaw formula remains unchanged.
Thus, in the double seesaw expression the former mass matrix $M_{S}$ ($A$ in the notation here)
gets replaced by an effective mass matrix
\begin{equation}\label{seesaw-like}
M_{S}^\text{eff} \equiv A - BC^{-1}B^T.
\end{equation}
This looks like the first term in a seesaw expansion, but the derivation
of Eq.~(\ref{modified-DS}) does not need the assumption of any hierarchy in
the matrix
\begin{equation}
M \equiv \begin{pmatrix}
A & B\\
B^T & C
\end{pmatrix}.
\end{equation}
The only condition is that $C$ is invertible. Therefore, singlet
fermions which do not couple directly to the $\sixt_F$ will
have an impact on the neutrino mass matrix $m_\nu^{DS}$.

The seesaw-like formula~(\ref{seesaw-like}) may offer more possibilities
to obtain interesting structures in $M_S$ and therefore in $m_\nu^{DS}$, in particular
when non-Abelian symmetries of $M$ are assumed (see section~\ref{ill2}).
If the symmetry $G_\text{hidden}$ of $M$ is Abelian, \textit{i.e.}\ $M$ is restricted by texture
zeros only, there could still be ``effective'' non-Abelian structures in $M_S^\text{eff}$. For example,
in case of a large hidden fermion sector, as could be motivated by string theory~\cite{Buchmueller,Ellis},
$M$ would have a large dimension of up to $\mathcal{O}(100)$, and therefore the elements of $M_S^\text{eff}$
would get many contributions. Since $M$ is
close to string/Planck scale, the Yukawa couplings could be universal (nearly equal
or with small spread).
If also the flavon VEVs determining $M$ are not completely random,
some of the elements of $M_S^\text{eff}$ could be approximately equal as is required
by the 2-3-permutation symmetry.
In this way specific structures in the $3\times 3$ effective mass
matrix $M_S^\text{eff}$ are possible without non-Abelian
symmetries in the hidden sector.

\section{CKM new physics}
\label{ckmphysics}

Up to now in our examples we discussed only
the Higgs ten-plet $\ten_H^u$ which has Higgs-doublet VEVs
$v_u\neq 0$ and $v_d=0$. This was enough for the
symmetry considerations regarding $m_D$, but in order to have quark mixing
we need $m_u\not\propto m_d$, which can be achieved by
introduction of scalar ten-plets $\ten_H^d$ with VEVs $v_u=0$ and $v_d\neq 0$.
Another possibility is
the introduction of additional scalars $\sixt_H'$ and $\sixt_H''$
giving $m_u\not\propto m_d$ via the effective operator
$(1/\Lambda) \sixt_F\sixt_F\sixt_H'\sixt_H''$~\cite{md1,md2,md3,md4}.
Here we will only discuss the scenario with additional ten-plets.

Generation of CKM mixing requires breaking of the basis-fixing symmetry
$G_\text{basis} = \zed_2 \times \zed_2$. In fact, $G_\text{basis}$ is already broken
spontaneously in the hidden sector by the flavon VEVs which generate
the matrix $M_S$. This breaking leads to quark mixing via higher order operators
\begin{equation}
\sixt_F \sixt_F \ten_H^d \frac{\one_{\chi ij} \one_{\chi kl}}{\Lambda^2}.
\label{basis-breaking}
\end{equation}
Here the singlet operator is built 
from the flavons of table~\ref{flavons}.
Invariance under $G_\text{aux}$ requires it
to be of second order. Note that it transforms non-trivially
under $G_\text{basis}$ and $G_\text{Yukawa}$. Hence, the operator
of Eq.~(\ref{basis-breaking}) gives corrections to the diagonal form of $m_d$
of the order $\langle\one_\chi\rangle^2/\Lambda^2\sim 10^{-2}$,
which are certainly too small to generate the experimentally observed $V_\text{CKM}$.
Therefore, we should introduce additional sources of $G_\text{basis}$ breaking. For
instance, breaking of $G_\text{basis}$ can be achieved by
introduction of
several ten-plets $\ten_H^d$ charged under $G_\text{basis}$ and $G_\text{Yukawa}$.
Instead, we can introduce only one $\ten_H^d$, being a singlet of $G_\text{basis}$,
which is connected by a symmetry $G_\text{conn,d}$ with additional flavons
$\one_{Y}'$ charged under $G_\text{basis}$ and $G_\text{Yukawa}$.
As connecting symmetry we can use
\begin{equation}
G_\text{conn,d}: \quad \ten_{H}^d \rightarrow \alpha \ten_{H}^d,\quad (\one_{Y}')^j \rightarrow \alpha^\ast (\one_{Y}')^j,
\end{equation}
where $\alpha$ is a root of unity and $j$ is a positive integer.
All elements of $m_d$ can then be generated by effective operators of the form
\begin{equation}
\sixt_F \sixt_F  \ten_{H}^d \left( \frac{\one_{Y}'}{\Lambda} \right)^j.
\end{equation}
The correct hierarchy of $m_d$ can be produced by appropriate values of
the VEVs and Yukawa-couplings of the different $\one_Y'$ as well as the powers $j$.
Elaboration of this new physics is beyond the scope of this paper.

An important feature of this scenario is that $V_\text{CKM}\neq \bone$ is
generated solely by new physics that determines the structure of $m_d$ and $m_\ell$. Thus, the
smallness of the 1-3 and 2-3 quark mixing angles is not related to the strong mass hierarchy
of the up-type quarks.
However, the size of the Cabibbo angle may still
be a result of the down-type quark mass hierarchy~\cite{Gatto,Cabibbo}.

Finally, to generate the correct mass hierarchy
in $m_\ell$, different from $m_d$, we have to introduce other representations for the Higgs fields.
For example, adding 45-plet Higgs fields $\underline{45}_H$ one can effectively
generate $m_\ell \not\propto m_d$ via the dimension-five operator~\cite{md1}
\begin{equation}\label{45}
\frac{1}{\Lambda} \sixt_F \sixt_F \ten_H \underline{45}_H.
\end{equation}
The problem is that inequality $m_\ell \neq m_d$ in general implies that
$U_\ell \neq U_d$ and thus spoils relation~(\ref{quark-lepton-mixing}).
A possible solution is that
the antisymmetric parts of $m_d$ and $m_\ell$ coming from the interaction~(\ref{45})
are chosen in such a way that still $U_\ell \approx U_d$, but $U_{\ell R} \neq U_{dR}$
and $\hat{m}_d \not\propto \hat{m}_\ell$~\cite{Antusch1,Antusch-Maurer}.

\section{Discussion and Conclusions}
\label{discussion}

The latest measurements of the lepton mixing parameters
are in good agreement with the relation $U_\text{PMNS} \approx V_\text{CKM}^{\dagger} U_X$. 
In this paper we have proposed a framework which allows to realize
such a relation. The framework is based on the double seesaw mechanism of neutrino mass generation, 
Grand Unification and the existence of a hidden sector with certain symmetries.

The framework provides another way to unification of the quark and lepton mixings.
The lepton mixing matrix has two contributions. The first one comes from the ``CKM new physics,'' which is common 
for quarks and leptons and associated to interactions that generate the Dirac mass matrices. 
The other one is the contribution from structures responsible for the
smallness of neutrino masses. 
These structures originate from the hidden sector.
In this framework the CKM physics can 
be disentangled to a large extent from the ``new neutrino physics.''
It allows to reconcile the strong mass hierarchy and small mixing of quarks with
the mild hierarchy of light neutrinos and large lepton mixing.

The key elements of the framework are the basis-fixing symmetry
and mediator fields which communicate information from the hidden sector to the visible one.
An appropriate basis-fixing symmetry is the Abelian $\zed_2 \times \zed_2$ for which the
symmetry basis coincides with the mass basis (in the first approximation).
The mediator fields have charges of the basis-fixing symmetry, as well as
charges of other symmetries in the hidden and/or visible sector.
The basis-fixing symmetry is consistent with
Abelian symmetries of the hidden sector which can lead
to particular structures of $M_X$, \textit{e.g.}\ a dominant
2-3-block.

To obtain $U_X$ of the required form,
non-Abelian symmetries in the hidden sector are required.
These non-Abelian symmetries $G_\text{hidden}$ (\textit{e.g.}\ a
2-3 permutation symmetry)
are broken in the portal interactions explicitly or spontaneously.   
Due to the lower scale of these interactions,
the symmetry breaking produces only small corrections to the case of exact symmetry under $G_\text{hidden}$.
The groups $G_\text{basis}$ and $G_\text{hidden}$ (and also $G_\text{Yukawa}$) can be embedded in
a bigger group $G_f$. In this way one can
realize the approach of residual symmetries such that
$G_f$ is broken down to $G_\text{hidden}$ in the
hidden sector and to $G_\text{basis}\times G_\text{Yukawa}$ in the portal
and visible sector interactions.

The mediator fields have to appear in the portal interaction and can be
$\overline{\sixt}_H$, or $\one_S$, or some new flavon fields associated to
$\overline{\sixt}_H$ or $\one_S$. This association needs an additional connecting symmetry
with respect to which the flavons and $\overline{\sixt}_H$ or $\one_S$ are charged.
For each of these possibilities we provide illustrative examples.

An extended hidden sector with more than three singlets $\one_S$ may
open up additional possibilities to explain features of
$U_X$. The additional fermionic singlets $\one_S'$ should not
participate directly in the portal interaction, otherwise screening
will be destroyed.
They can, however, couple with the three $\one_S$, thus modifying the matrix
$M_S$, and in this way can $\textit{e.g.}$\ lead to an effective 2-3-permutation symmetry in $M_S$.

New physics and new symmetries are involved in
the generation of the CKM mixing and the mass hierarchies of
quarks and leptons. 
In the first approximation SO(10) allows to
disentangle mixing and masses in a very simple way:
if a single $\ten_H$ gives the dominant contribution to the masses,
no mixing is produced.
So, we are forced to consider at least partly the
CKM physics. In order to disentangle the generation
of $m_u$ and $m_D$ from $m_d$ and $m_\ell$, we introduced
two Higgs ten-plets $\ten_H^u$ and $\ten_H^d$ with
identical charges under $G_\text{basis}$.
Additional flavons, charged under $G_\text{basis}$ and $G_\text{Yukawa}$
and connected to the field $\ten_H^d$ by a symmetry
$G_\text{conn,d}$, allow to generate a non-diagonal $m_d\propto m_\ell$.
The difference between $m_d$ and $m_\ell$ should be
obtained in such a way that the relation~(\ref{quark-lepton-mixing})
is not destroyed.

A crucial question is how to test the proposed framework.
Some possibilities are:
\begin{enumerate}
 \item Further more accurate measurements of the leptonic 1-3
and 2-3 mixing, and especially the determination of the quadrant of
the 2-3 mixing are important. The second quadrant for the 2-3 mixing angle
would disfavour the framework.
 \item Specific realizations of the framework may lead to certain predictions for
the Dirac as well as Majorana CP phases.
 \item The neutrino masses are generated at high scales
which are not accessible to direct experimental studies.
Therefore, one should not expect to see any
new physics at the LHC associated directly
to neutrino mass generation. Observation of such physics
would probably exclude the framework.
 \item Indirect support of the approach can be provided by the observation
of proton decay (which will be in favour of Grand Unification).
 \item A connection to leptogenesis and inflation may give another test of the high scale physics.
 \item Discoveries of other possible manifestations of the hidden sector
are important. That includes identification of the dark matter and
the discovery of sterile neutrinos.
\end{enumerate}

\paragraph{Acknowledgements:} P.O.L.\ wants to thank all colleagues
at the MPIK Heidelberg---in particular
Stefan Br\"unner,
Miguel Campos,
Julia Haser,
Kher Sham Lim,
Yusuke Shimizu,
Anne Wegmann and
Xunjie Xu---for the excellent and encouraging working atmosphere.


\begin{thebibliography}{99}
%

%
\bibitem{Giunti-Tanimoto}
  C.~Giunti and M.~Tanimoto,
  \textit{Deviation of neutrino mixing from bimaximal},
  Phys.\ Rev.\ D {\bf 66} (2002) 053013
  [hep-ph/0207096].
%
\bibitem{Minakata-Smirnov}
  H.~Minakata and A.~Y.~Smirnov,
  \textit{Neutrino mixing and quark-lepton complementarity},
  Phys.\ Rev.\ D {\bf 70} (2004) 073009
  [hep-ph/0405088].


%
\bibitem{Xing}
  Z.~z.~Xing,
  \textit{Nontrivial correlation between the CKM and MNS matrices},
  Phys.\ Lett.\ B {\bf 618} (2005) 141
  [hep-ph/0503200].
%
\bibitem{Harada}
  J.~Harada,
  \textit{Neutrino mixing and CP violation from Dirac-Majorana bimaximal mixture and quark-lepton unification},
  Europhys.\ Lett.\  {\bf 75} (2006) 248
  [hep-ph/0512294].
%
\bibitem{Antusch1}
  S.~Antusch, S.~F.~King and R.~N.~Mohapatra,
  \textit{Quark-lepton complementarity in unified theories},
  Phys.\ Lett.\ B {\bf 618} (2005) 150
  [hep-ph/0504007].
%
\bibitem{Antusch2}
  S.~Antusch and S.~F.~King,
  \textit{Charged lepton corrections to neutrino mixing angles and CP phases revisited},
  Phys.\ Lett.\ B {\bf 631} (2005) 42
  [hep-ph/0508044].
%
\bibitem{Farzan}
  Y.~Farzan and A.~Y.~Smirnov,
  \textit{Leptonic CP violation: Zero, maximal or between the two extremes},
  JHEP {\bf 0701} (2007) 059
  [hep-ph/0610337].
%
\bibitem{Picariello}
  B.~C.~Chauhan, M.~Picariello, J.~Pulido and E.~Torrente-Lujan,
  \textit{Quark-lepton complementarity, neutrino and standard model data predict $\theta^\text{PMNS}_{13} = (9^{+1}_{-2})^\circ$},
  Eur.\ Phys.\ J.\ C {\bf 50} (2007) 573
  [hep-ph/0605032].


%
\bibitem{bimaximal1}
  F.~Vissani,
  \textit{A Study of the scenario with nearly degenerate Majorana neutrinos},
  hep-ph/9708483.
%
\bibitem{bimaximal2}
  V.~D.~Barger, S.~Pakvasa, T.~J.~Weiler and K.~Whisnant,
  \textit{Bimaximal mixing of three neutrinos},
  Phys.\ Lett.\ B {\bf 437} (1998) 107
  [hep-ph/9806387].


%
\bibitem{King}
  S.~F.~King,
  \textit{Tri-bimaximal-Cabibbo Mixing},
  Phys.\ Lett.\ B {\bf 718} (2012) 136
  [arXiv:1205.0506 [hep-ph]].


%
\bibitem{HPS-TBM}
  P.~F.~Harrison, D.~H.~Perkins and W.~G.~Scott,
  \textit{Tri-bimaximal mixing and the neutrino oscillation data},
  Phys.\ Lett.\ B {\bf 530} (2002) 167
  [hep-ph/0202074].


%
\bibitem{golden-ratio}
  Y.~Kajiyama, M.~Raidal and A.~Strumia,
  \textit{The Golden ratio prediction for the solar neutrino mixing},
  Phys.\ Rev.\ D {\bf 76} (2007) 117301
  [arXiv:0705.4559 [hep-ph]].


%
\bibitem{DoubleChooz}
  Y.~Abe {\it et al.}  [Double Chooz Collaboration],
  \textit{Improved measurements of the neutrino mixing angle $\theta_{13}$ with the Double Chooz detector},
  JHEP {\bf 1410} (2014) 086
  [JHEP {\bf 1502} (2015) 074]
  [arXiv:1406.7763 [hep-ex]].
%
\bibitem{DayaBay}
  B.~Z.~Hu [for the Daya Bay Collaboration],
  \textit{Recent Results from Daya Bay Reactor Neutrino Experiment},
  arXiv:1505.03641 [hep-ex].
%
\bibitem{RENO}
  J.~K.~Ahn {\it et al.}  [RENO Collaboration],
  \textit{Observation of Reactor Electron Antineutrino Disappearance in the RENO Experiment},
  Phys.\ Rev.\ Lett.\  {\bf 108} (2012) 191802
  [arXiv:1204.0626 [hep-ex]].




%
\bibitem{PDG}
  K.~A.~Olive {\it et al.} [Particle Data Group],
  \textit{Review of Particle Physics},
  Chin.\ Phys.\ C {\bf 38} (2014) 090001.


%
\bibitem{Kopp}
  J.~Kopp, P.~A.~N.~Machado, M.~Maltoni and T.~Schwetz,
  \textit{Sterile Neutrino Oscillations: The Global Picture},
  JHEP {\bf 1305} (2013) 050
  [arXiv:1303.3011 [hep-ph]].


\bibitem{deHolanda}
  P.~C.~de Holanda and A.~Y.~Smirnov,
  \textit{Solar neutrino spectrum, sterile neutrinos and additional radiation in the Universe},
  Phys.\ Rev.\ D {\bf 83} (2011) 113011
  [arXiv:1012.5627 [hep-ph]].


%
\bibitem{LSND}
  A.~Aguilar-Arevalo {\it et al.}  [LSND Collaboration],
  \textit{Evidence for neutrino oscillations from the observation of anti-neutrino(electron) appearance in a anti-neutrino(muon) beam},
  Phys.\ Rev.\ D {\bf 64} (2001) 112007
  [hep-ex/0104049].


%
\bibitem{MiniBooNE}
  A.~A.~Aguilar-Arevalo {\it et al.}  [MiniBooNE Collaboration],
  \textit{Improved Search for $\bar \nu_\mu \rightarrow \bar \nu_e$ Oscillations in the MiniBooNE Experiment},
  Phys.\ Rev.\ Lett.\  {\bf 110} (2013) 161801
  [arXiv:1207.4809 [hep-ex], arXiv:1303.2588 [hep-ex]].


%
\bibitem{reactor-anomaly}
  G.~Mention, M.~Fechner, T.~Lasserre, T.~A.~Mueller, D.~Lhuillier, M.~Cribier and A.~Letourneau,
  \textit{The Reactor Antineutrino Anomaly},
  Phys.\ Rev.\ D {\bf 83} (2011) 073006
  [arXiv:1101.2755 [hep-ex]].


%
\bibitem{Gallex1}
  W.~Hampel {\it et al.}  [GALLEX Collaboration],
  \textit{Final results of the Cr-51 neutrino source experiments in GALLEX},
  Phys.\ Lett.\ B {\bf 420} (1998) 114.
%
\bibitem{Gallex2}
  F.~Kaether, W.~Hampel, G.~Heusser, J.~Kiko and T.~Kirsten,
  \textit{Reanalysis of the GALLEX solar neutrino flux and source experiments},
  Phys.\ Lett.\ B {\bf 685} (2010) 47
  [arXiv:1001.2731 [hep-ex]].
%
\bibitem{SAGE1}
  J.~N.~Abdurashitov {\it et al.}  [SAGE Collaboration],
  \textit{Measurement of the response of the Russian-American gallium experiment to neutrinos from a Cr-51 source},
  Phys.\ Rev.\ C {\bf 59} (1999) 2246
  [hep-ph/9803418].
%
\bibitem{SAGE2}
  J.~N.~Abdurashitov {\it et al.},
  \textit{Measurement of the response of a Ga solar neutrino experiment to neutrinos from an Ar-37 source},
  Phys.\ Rev.\ C {\bf 73} (2006) 045805
  [nucl-ex/0512041].


%
\bibitem{WDM}
  S.~Dodelson and L.~M.~Widrow,
  \textit{Sterile-neutrinos as dark matter},
  Phys.\ Rev.\ Lett.\  {\bf 72} (1994) 17
  [hep-ph/9303287].


%
\bibitem{Buchmueller}
  W.~Buchm\"uller, K.~Hamaguchi, O.~Lebedev, S.~Ramos-Sanchez and M.~Ratz,
  \textit{Seesaw neutrinos from the heterotic string},
  Phys.\ Rev.\ Lett.\  {\bf 99} (2007) 021601
  [hep-ph/0703078].


%
\bibitem{Wolfenstein}
  L.~Wolfenstein,
  \textit{Parametrization of the Kobayashi-Maskawa Matrix},
  Phys.\ Rev.\ Lett.\  {\bf 51} (1983) 1945.


%
\bibitem{Schwetz}
  M.~C.~Gonzalez-Garcia, M.~Maltoni and T.~Schwetz,
  \textit{Updated fit to three neutrino mixing: status of leptonic CP violation},
  JHEP {\bf 1411} (2014) 052
  [arXiv:1409.5439 [hep-ph]].

\bibitem{NuFIT}
  M.~C.~Gonzalez-Garcia, M.~Maltoni, and T.~Schwetz,
  \textit{NuFIT 2.0} (2014),
  http://www.nu-fit.org.


%
\bibitem{Raidal}
  M.~Raidal,
  \textit{Relation between the neutrino and quark mixing angles and grand unification},
  Phys.\ Rev.\ Lett.\  {\bf 93} (2004) 161801
  [hep-ph/0404046].


%
\bibitem{Seesaw1}
P.~Minkowski,
\textit{$\mu\rightarrow e\gamma$ at a rate of one out of $10^9$ muon decays?},
Phys.\ Lett.\ B {\bf 67} (1977) 421.

\bibitem{Seesaw2}
T.~Yanagida, in \textit{Proceedings of the Workshop on Unified Theory and Baryon
Number in the Universe},
O.~Sawata and A.~Sugamoto eds.,
KEK report {\bf 79-18}, Tsukuba, Japan (1979).

\bibitem{Seesaw3}
S.~L.~Glashow, in \textit{Quarks and Leptons, Proceedings of the Advanced Study Institute
(Carg\`ese, Corsica, 1979)},
J.-L.~Basdevant \textit{et al.} eds.,
Plenum, New York (1981).

\bibitem{Seesaw4}
M.~Gell-Mann, P.~Ramond and R.~Slansky,
\textit{Complex spinors and unified theories}, in \textit{Supergravity},
D.~Z.~Freedman and F.~van~Nieuwenhuizen eds.,
North Holland, Amsterdam (1979).

\bibitem{Seesaw5}
R.~N.~Mohapatra and G.~Senjanovic,
\textit{Neutrino mass and spontaneous parity violation},
Phys.\ Rev.\ Lett.\  {\bf 44} (1980) 912.


%
\bibitem{GUT1}
  J.~C.~Pati and A.~Salam,
  \textit{Unified Lepton-Hadron Symmetry and a Gauge Theory of the Basic Interactions},
  Phys.\ Rev.\ D {\bf 8} (1973) 1240.
%
\bibitem{GUT2}
  H.~Georgi and S.~L.~Glashow,
  \textit{Unity of All Elementary Particle Forces},
  Phys.\ Rev.\ Lett.\  {\bf 32} (1974) 438.


%
\bibitem{SO10-1}
  H.~Georgi,
  \textit{The State of the Art---Gauge Theories},
  AIP Conf.\ Proc.\  {\bf 23} (1975) 575.
%
\bibitem{SO10-2}
  H.~Fritzsch and P.~Minkowski,
  \textit{Unified Interactions of Leptons and Hadrons},
  Annals Phys.\  {\bf 93} (1975) 193.


%
\bibitem{Sluka}
  S.~Antusch, C.~Gross, V.~Maurer and C.~Sluka,
  \textit{$\theta^\mathrm{PMNS}_{13} = \theta_C / \sqrt2$ from GUTs},
  Nucl.\ Phys.\ B {\bf 866} (2013) 255
  [arXiv:1205.1051 [hep-ph]].


%
\bibitem{double-seesaw1}
  R.~N.~Mohapatra,
  \textit{Mechanism for Understanding Small Neutrino Mass in Superstring Theories},
  Phys.\ Rev.\ Lett.\ {\bf 56} (1986) 561.
%
\bibitem{double-seesaw2}
  R.~N.~Mohapatra and J.~W.~F.~Valle,
  \textit{Neutrino Mass and Baryon Number Nonconservation in Superstring Models},
  Phys.\ Rev.\ D {\bf 34} (1986) 1642.


%
\bibitem{linear-seesaw}
  S.~M.~Barr,
  \textit{A Different seesaw formula for neutrino masses},
  Phys.\ Rev.\ Lett.\  {\bf 92} (2004) 101601
  [hep-ph/0309152].


%
\bibitem{Akhmedov}
  E.~K.~Akhmedov, M.~Frigerio and A.~Y.~Smirnov,
  \textit{Probing the seesaw mechanism with neutrino data and leptogenesis},
  JHEP {\bf 0309} (2003) 021
  [hep-ph/0305322].


%
\bibitem{ten-u-d-1}
  W.~Buchm\"uller and D.~Wyler,
  \textit{CP violation, neutrino mixing and the baryon asymmetry},
  Phys.\ Lett.\ B {\bf 521} (2001) 291
  [hep-ph/0108216].
%
\bibitem{ten-u-d-2}
  A.~Masiero, S.~K.~Vempati and O.~Vives,
  \textit{Seesaw and lepton flavor violation in SUSY SO(10)},
  Nucl.\ Phys.\ B {\bf 649} (2003) 189
  [hep-ph/0209303].




%
\bibitem{screening1}
  A.~Y.~Smirnov,
  \textit{Seesaw enhancement of lepton mixing},
  Phys.\ Rev.\ D {\bf 48} (1993) 3264
  [hep-ph/9304205].
%
\bibitem{screening2}
  M.~Lindner, M.~A.~Schmidt and A.~Y.~Smirnov,
  \textit{Screening of Dirac flavor structure in the seesaw and neutrino mixing},
  JHEP {\bf 0507} (2005) 048
  [hep-ph/0505067].


%
\bibitem{lam1}
  C.~S.~Lam,
  \textit{Determining Horizontal Symmetry from Neutrino Mixing},
  Phys.\ Rev.\ Lett.\  {\bf 101} (2008) 121602
  [arXiv:0804.2622 [hep-ph]].
%
\bibitem{lam2}
  C.~S.~Lam,
  \textit{The Unique Horizontal Symmetry of Leptons},
  Phys.\ Rev.\ D {\bf 78} (2008) 073015
  [arXiv:0809.1185 [hep-ph]].
%
\bibitem{grimus}
  W.~Grimus, L.~Lavoura and P.~O.~Ludl,
  \textit{Is $S_4$ the horizontal symmetry of tri-bimaximal lepton mixing?},
  J.\ Phys.\ G {\bf 36} (2009) 115007
  [arXiv:0906.2689 [hep-ph]].
%
\bibitem{toorop}
  R.~d.~A.~Toorop, F.~Feruglio and C.~Hagedorn,
  \textit{Discrete Flavour Symmetries in Light of T2K},
  Phys.\ Lett.\ B {\bf 703} (2011) 447
  [arXiv:1107.3486 [hep-ph]].
%
\bibitem{hernandez}
  D.~Hernandez and A.~Y.~Smirnov,
  \textit{Lepton mixing and discrete symmetries},
  Phys.\ Rev.\ D {\bf 86} (2012) 053014
  [arXiv:1204.0445 [hep-ph]].
%
\bibitem{holthausen}
  M.~Holthausen, K.~S.~Lim and M.~Lindner,
  \textit{Lepton Mixing Patterns from a Scan of Finite Discrete Groups},
  Phys.\ Lett.\ B {\bf 721} (2013) 61
  [arXiv:1212.2411 [hep-ph]].
%
\bibitem{fonseca}
  R.~M.~Fonseca and W.~Grimus,
  \textit{Classification of lepton mixing matrices from finite residual symmetries},
  JHEP {\bf 1409} (2014) 033
  [arXiv:1405.3678 [hep-ph]].




%
\bibitem{Hagedorn}
  C.~Hagedorn, M.~A.~Schmidt and A.~Y.~Smirnov,
  \textit{Lepton Mixing and Cancellation of the Dirac Mass Hierarchy in SO(10) GUTs with Flavor Symmetries $T_7$ and $\Sigma(81)$},
  Phys.\ Rev.\ D {\bf 79} (2009) 036002
  [arXiv:0811.2955 [hep-ph]].



%
\bibitem{E6-1}
  F.~Gursey, P.~Ramond and P.~Sikivie,
  \textit{A Universal Gauge Theory Model Based on E6},
  Phys.\ Lett.\ B {\bf 60} (1976) 177.
%
\bibitem{E6-2}
  Y.~Achiman and B.~Stech,
  \textit{Quark Lepton Symmetry and Mass Scales in an E6 Unified Gauge Model},
  Phys.\ Lett.\ B {\bf 77} (1978) 389.
%
\bibitem{E6-3}
  Q.~Shafi,
  \textit{E(6) as a Unifying Gauge Symmetry},
  Phys.\ Lett.\ B {\bf 79} (1978) 301.
%
\bibitem{E6-4}
  R.~Barbieri, D.~V.~Nanopoulos and A.~Masiero,
  \textit{Hierarchical Fermion Masses in E6},
  Phys.\ Lett.\ B {\bf 104} (1981) 194.


%
\bibitem{dihedral}
  A.~Blum, C.~Hagedorn and M.~Lindner,
  \textit{Fermion Masses and Mixings from Dihedral Flavor Symmetries with Preserved Subgroups},
  Phys.\ Rev.\ D {\bf 77} (2008) 076004
  [arXiv:0709.3450 [hep-ph]].


%
\bibitem{Altarelli-Feruglio}
  G.~Altarelli and F.~Feruglio,
  \textit{Discrete Flavor Symmetries and Models of Neutrino Mixing},
  Rev.\ Mod.\ Phys.\  {\bf 82} (2010) 2701
  [arXiv:1002.0211 [hep-ph]].
%
\bibitem{Holthausen}
  M.~Holthausen and M.~A.~Schmidt,
  \textit{Natural Vacuum Alignment from Group Theory: The Minimal Case},
  JHEP {\bf 1201} (2012) 126
  [arXiv:1111.1730 [hep-ph]].


%
\bibitem{GAP}
  \textit{GAP - Groups, Algorithms, Programming -
  a System for Computational Discrete Algebra}. www.gap-system.org.


%
\bibitem{Bovier}
  A.~Bovier, M.~L\"uling and D.~Wyler,
  \textit{Finite Subgroups of SU(3)},
  J.\ Math.\ Phys.\  {\bf 22} (1981) 1543.
%
\bibitem{Luhn}
  C.~Luhn, S.~Nasri and P.~Ramond,
  \textit{The Flavor group $\Delta(3n^2)$},
  J.\ Math.\ Phys.\  {\bf 48} (2007) 073501
  [hep-th/0701188].


%
\bibitem{Grimus-Lavoura}
  W.~Grimus and L.~Lavoura,
  \textit{The Seesaw mechanism at arbitrary order: Disentangling the small scale from the large scale},
  JHEP {\bf 0011} (2000) 042
  [hep-ph/0008179].


%
\bibitem{Grimus-Joshipura}
  W.~Grimus, A.~S.~Joshipura, L.~Lavoura and M.~Tanimoto,
  \textit{Symmetry realization of texture zeros},
  Eur.\ Phys.\ J.\ C {\bf 36} (2004) 227
  [hep-ph/0405016].


%
\bibitem{Ellis}
  J.~R.~Ellis and O.~Lebedev,
  \textit{The Seesaw with many right-handed neutrinos},
  Phys.\ Lett.\ B {\bf 653} (2007) 411
  [arXiv:0707.3419 [hep-ph]].


%
\bibitem{md1}
  K.~S.~Babu, J.~C.~Pati and F.~Wilczek,
  \textit{Fermion masses, neutrino oscillations, and proton decay in the light of Super-Kamiokande},
  Nucl.\ Phys.\ B {\bf 566} (2000) 33
  [hep-ph/9812538].
%
\bibitem{md2}
  C.~H.~Albright and S.~M.~Barr,
  \textit{Predicting quark and lepton masses and mixings},
  Phys.\ Lett.\ B {\bf 452} (1999) 287
  [hep-ph/9901318].
%
\bibitem{md3}
  C.~H.~Albright and S.~M.~Barr,
  \textit{Explicit SO(10) supersymmetric grand unified model for the Higgs and Yukawa sectors},
  Phys.\ Rev.\ Lett.\  {\bf 85} (2000) 244
  [hep-ph/0002155].
%
\bibitem{md4}
  C.~H.~Albright and S.~M.~Barr,
  \textit{Construction of a minimal Higgs SO(10) SUSY GUT model},
  Phys.\ Rev.\ D {\bf 62} (2000) 093008
  [hep-ph/0003251].


%
\bibitem{Gatto}
  R.~Gatto, G.~Sartori and M.~Tonin,
  \textit{Weak self-masses, Cabibbo angle, and broken $SU_2 \times SU_2$},
  Phys.\ Lett.\ \textbf{28B} (1968) 128.
%
\bibitem{Cabibbo}
  N.~Cabibbo and L.~Maiani,
  \textit{Dynamical interrelations of weak, electromagnetic and strong
  interactions and the value of $\theta$},
  Phys.\ Lett.\ \textbf{28B} (1968) 131.


%
\bibitem{Antusch-Maurer}
  S.~Antusch and V.~Maurer,
  \textit{Large neutrino mixing angle $\theta_{13}^{MNS}$ and quark-lepton mass ratios in unified flavour models},
  Phys.\ Rev.\ D {\bf 84} (2011) 117301
  [arXiv:1107.3728 [hep-ph]].






\end{thebibliography}
\end{document}